%% file: main.tex
\documentclass[10pt,journal,compsoc]{IEEEtran}
\ifCLASSOPTIONcompsoc
  \usepackage[nocompress]{cite}
\else
  \usepackage{cite}
\fi
\ifCLASSINFOpdf
\else
\fi
\newcommand{\ana}[1]{}
\newcommand{\eb}[1]{}
\newcommand{\ms}[1]{}

\newcommand{\rev}[1]{\textcolor{black}{#1}}

\newcommand{\revMR}[1]{\textcolor{black}{#1}}
\newcommand{\revMI}[1]{\textcolor{black}{#1}}

\usepackage{wrapfig}
\usepackage{enumitem}
\usepackage{soul}
\usepackage{tikz}
\usepackage{tikz-cd}
\usepackage[hidelinks]{hyperref}

\usepackage{adjustbox}

\begin{document}
%
\title{Eliciting Model Steering Interactions from Users via Data and Visual Design Probes}


%
%
%
%

\author{Anamaria Crisan,
        Maddie Shang,
        and~Eric Brochu
\IEEEcompsocitemizethanks{\IEEEcompsocthanksitem A. Crisan is with Tableau Research, Seattle, USA, 98103
\IEEEcompsocthanksitem M. Shang and E. Brochu are with Tableau, Vancouver, CA, V6B 1A6.\protect\\
E-mail: \{acrisan, mshang,e brochu\}@tableau.com
}

\thanks{Manuscript received April 19, 2005; revised August 26, 2015.}}

%
%

\markboth{Journal of \LaTeX\ Class Files,~Vol.~14, No.~8, August~2015}%
{Shell \MakeLowercase{\textit{et al.}}: Bare Demo of IEEEtran.cls for Computer Society Journals}
%



\IEEEtitleabstractindextext{%
\revMR{Visual and interactive machine learning systems (IML) are becoming ubiquitous as they empower individuals with varied machine learning expertise to analyze data. However, it remains complex to align interactions with visual marks to a user's intent for steering machine learning models. }
We explore using data and visual design probes to elicit users' desired interactions to steer ML models via visual encodings within IML interfaces. We conducted an elicitation study with 20  data analysts with varying expertise in ML. We summarize our findings as pairs of target-interaction, which we compare to prior systems to assess the utility of the probes. We additionally surfaced insights about factors influencing how and why participants chose to interact with visual encodings, including refraining from interacting. \revMR{Finally, we reflect on the value of gathering such formative empirical evidence via data and visual design probes ahead of developing IML prototypes.}  


\begin{IEEEkeywords}
Design Probes, Interactive Machine Learning, Model Steering, Semantic Interactions
\end{IEEEkeywords}}

\maketitle

\IEEEdisplaynontitleabstractindextext

%
\IEEEpeerreviewmaketitle

\IEEEraisesectionheading{\section{Introduction}\label{sec:introduction}}

\input{tex/intro}
\input{tex/related_work}

\input{tex/elicitation_study}
\input{tex/full_study}
\input{tex/discussion_conclusion}

\bibliographystyle{IEEEtran}
\scriptsize
\bibliography{main.bib} 

%

\vspace{-10mm}
\begin{IEEEbiographynophoto}{Anamaria Crisan}
is a Lead Research Scientist at
Tableau. She holds a PhD in Computer Science from the University of British Columbia. Her research explores the intersection of data science and data visualization; her prior work has emphasized biomedical applications. 
\end{IEEEbiographynophoto}

\vspace{-10mm}
\begin{IEEEbiographynophoto}{Maddie Shang}
Maddie Shang is a Sr. Research Engineer and Team Lead of OpenMined RecSys. She has degree in Math from UWaterloo. Her open-source work in Federated Learning and Privacy was funded by PyTorch. Her interests include HCI via NLP and visual encodings, AI Ethics and Reinforcement Learning. She was named as one of 100 Brilliant Women in AI Ethics in 2022.
\end{IEEEbiographynophoto}


\vspace{-10mm}
\begin{IEEEbiographynophoto}{Eric Brochu}
is a Software Engineering Architect at Tableau. He completed his PhD at the University of British Columbia in 2010, working on human-in-the-loop Bayesian optimization. He currently leads Tableau's Machine Learning Engineering team.
\end{IEEEbiographynophoto}




\end{document}

%% file: tex/intro.tex
\rev{Visual and Interactive machine learning (IML) systems are powerful interfaces for incorporating humans into the machine learning loop~\cite{collins:2018:guidance,sacha:2017:wywc,SPERRLE:coadative:2021,Sperrle:HCE_STAR:2021}. \revMR{However, designing visual encodings and interactions for steering machine learning models is a complex problem~\cite{Boukhelifa:eval_attribution:2020,dudley:UI_IML:2018}. One source of this challenge is the tension between model and data-centric priorities~\cite{Combemale:DACI:2021,Subramonyam:DACI:2021,Sambasivan:DataWork:2021} that require different interaction modalities, for example using control panels or direct manipulation of marks~\cite{Ender:beyond_controlpanels:2013}. Another is a broad user base with diverse expertise in machine learning (ML) and data science (DS), but who regularly employ these systems~\cite{Qian:2018:MTeaching,Swati:design_nonexperts:2021}. Finally,  there are challenges toward aligning human intent with ML and data interactions~\cite{Yang:HAI_hard:2020,dudley:UI_IML:2018}.}}

\revMR{We explore the use of data and visual design probes as a way to gather formative empirical evidence to align visual IML systems interactions with users' anticipated outcomes and intents.} \rev{Probes are an important component of pre-design formative research capable of co-creating user experiences with impacted stakeholders and surfacing problems ahead of development effort~\cite{Sanders:Probes:2014}.} \revMR{They can be useful instruments for assessing the \textit{semantic distance}~\cite{Hutchins1985DirectMI} between a user's actions and the meaning behind them - for example, whether changing the position of a mark suffices a user's intent of modifying its class label.}
\rev{While probes exist in a variety of forms and are widely used in human computer interaction (HCI) research~\cite{Sanders:Probes:2014,Graham:probes:2008}, including prior IML applications,~\cite{Yang:HAI_hard:2020, Subramonyam:Prtotyping:2021} they are under-explored in visualization research. \revMR{Instead, visualization research has tended to emphasize the development and evaluation of prototypes, which can result in cementing design choices ahead of sufficient formative input from end users~\cite{Sanders:Probes:2014}. Prototypes can also suffer from the \textit{attribution problem} and can lack ``\textit{observed results are loosely attributed to the system as a whole, but without 
accurate explanations or insights as to what component(s) played a bigger role  to achieve those results''}~\cite{Boukhelifa:eval_attribution:2020}}}

\revMR{Here, we developed a set of data and visual design probes that we used to conduct an elicitation study with twenty participants that routinely analyze data but vary in their ML and DS expertise. The goals of our study are threefold. First, we seek to elicit participants' intents coupled with direct manipulations of visual marks to update an ML model.} We deliberately place a lower emphasis on other features of a visual IML system and focus solely on single visual encodings to mitigate attribution issues~\cite{Boukhelifa:eval_attribution:2020}. \revMR{Second, we explore to what extent and in what ways the interactions differ in accordance with ML/DS expertise; while we can make some basic assumptions the goal is to gather data in lieu of making design decisions on instinct. Finally, we reflect on how the empirical evidence we gather can be used to prioritize the elements of visual IML systems.} 

\noindent\textbf{Collectively, we present three core contributions}:

\vspace{-0.5mm}
\begin{itemize}
\setlength{\itemsep}{0.5pt}
    \item  The design of data and visual design probes for examining direct manipulations of visual encodings
    \item A set of target-interaction pairs that describe both \textit{what} data interactions act upon and \textit{how}
    \item  The results of an elicitation study that explores factors influencing participants' interactions
\end{itemize}

\vspace{-0.5mm}
\revMR{These contributions present opportunities for expanding the visualization research toolbox of formative methodologies for delineating an interaction design space that aligns with users' needs and expectations.}

%% file: tex/related_work.tex
\section{Related Work}\label{sec:relatedwork}
We summarize pertinent research from visualization, human computer interaction, and machine learning research.

\vspace{-2mm}
\subsection{Frameworks for Human-AI/ML Interaction}
\rev{Theoretical frameworks for paradigms for human-AI/ML interaction inform and influence the development of techniques for visual and interactive machine learning~\cite{Lee2019AHP,Zanzotto:HAI:2019,Crisan:2021:automl,dellermann:future_collab:2021,Ben:2020:HCAI,Gil:HGML:2019}. Within the visualization community, these frameworks are rooted in earlier articulations of interactions more generally.} Ceneda~\textit{et al.}~\cite{Ceneda:2017:guidance} presented a framework that incorporated human guidance into the visualization processes proposed by van Wikj's~\cite{Wijk:2005} model of interactions. Building on their processes is the work by Sperrle~\textit{et al.} ~\cite{SPERRLE:coadative:2021}, who present a framework for a co-adaptive process for people to both learn from a model and incrementally guide it toward a better understanding of the human's analytic intents over time. They describe action-reaction pairs that complement the findings of the target-interaction pairs from the elicitation study.  Sacha~\textit{et al.} ~\cite{sacha:2017:wywc} and Collins~\textit{et al.} ~\cite{collins:2018:guidance} proposed frameworks to address limitations of incorporating human guidance via visual analytics processes. They delineate various points and methods for human intervention, including semantic interactions, across stages of model building.  Beyond this work, others have recently explored how an algebraic visualization design process (AVD)~\cite{Kindlman:2014} can help surface the side-effects of ML model updates on visual representations~\cite{Crisan:2021:automl,McNutt:2020:CHI}. Dudely~\textit{et al.} ~\cite{dudley:UI_IML:2018} place visualization tools in the context of broader user interface affordances and describe how they can benefit end-users with varying degrees of ML experts

\rev{The development of our visual design probes is influenced by these prior frameworks for human-AI/ML interaction. We were especially influenced by Sacha's~\cite{sacha:2017:wywc} framework of ``what you see is what you can change'', to enable direct manipulations with data encoding marks.}

\vspace{-2mm}
\subsection{Techniques for Interactive Model Steering}
Direct manipulations for user interface elements have long been a component in steering models of IML systems~\cite{ben:DM:1983,Jiang2019RecentRA,dudley:UI_IML:2018}. A common approach is to allow users to change the model's parameters by directly changing the values of input widgets (i.e., change a number in an input box or slider). Although common, this approach has limitations)~\cite{Endert2017,Ender:beyond_controlpanels:2013}. Input widgets are `model-centric knobs' that can be overwhelm users and be ineffective\cite{dudley:UI_IML:2018,ramos2020interactive,Holstein:practitioner_fairness:2019}. 

\rev{Semantic Interactions are an attractive alternative to input widgets. Semantic interactions shield end-users from the complexity of the model while using a visual metaphor that interprets interactions with visual encodings to the underlying model parameters~\cite{Endert:2012:SemanticOG}. Moreover, these interactions are incorporated into model learning and provide feedback to end-users through the visual metaphor.} Endert~\textit{et al.} demonstrated the utility of these semantic interactions with the ForceSPIRE~\cite{Endert:2012:SemanticOG} system, with further extension by Bradel~\textit{et al.} to develop StarSPIRE~\cite{Bradel:Starspire:1995}. Semantic interaction techniques have also been applied to standard text modeling algorithms, as evidenced by iVisCluster~\cite{Lee:iVisClustering:2012} and Utopian~\cite{Choo:Utopian:2013}. While this early work focused on spatial metaphors, others have shown their limitations~\cite{Demiral:2014} and have explored complementary and alternative approaches.

Other research has explored direct manipulations with data via visual encodings but does not explicitly refer to themselves as semantic interactions. Saket~\textit{et al.}~\cite{Saket:2020:Direct_Manipulation,Saket:visdemo:2017,Hartmann:SensorDM:2007} have explored possible interactions with different encoding types; we leverage their ideas and approach in our design probes. Other systems explore visual encodings more coarsely than~\cite{Saket:2020:Direct_Manipulation,Saket:visdemo:2017,Hartmann:SensorDM:2007}, but prioritize the development of techniques for model steering.  Brown~\textit{et al.}~\cite{Brown:2012} demonstrates how interactions with spatialized data can learn and refine and update a distance function. Similarly, Podium learns a ranking function through users' interactions~\cite{wall:podium:2018}.  More recent work on collaborative semantic inference~\cite{Gehrmann:MLhooks:2020} uses MLhooks to capture interactions and update a deep learning model. Finally, there exist techniques that more directly scrutinize the effects of data changes, and not just parameter or function modifications, on the model changes. Systems like iCluster~\cite{Drucker:iCluster:2011} and VIAL~\cite{Bernard:VIAL:2018} demonstrate that visual clustering and labeling of data can outperform online learning alone~\cite{Bernard:2018}. The Chameleon~\cite{hohman:data_iteration:2020} prototype enables ML practitioners to visualize the effects of their data changes on the model's features and performance.

\rev{Here, we list only a subset of the many techniques that exist for incorporating human actions to steer ML models via visual and interactive interactions. We refer the reader to more recent systematic reviews for in-depth and comprehensive overview\cite{Endert2017,Sacha:ReviewDIM:2017,Yuan:survey:2021,Bae:iClustering:2020}. To reflect on the utility of our approach, we compare the interactions we elicit design probes to existing and varied techniques for human-AI/ML interaction and model steering.}

\vspace{-3mm}
\subsection{Design Probes for Interactive Machine Learning}\label{background-probes}
\rev{Design probes, generative toolkits, and prototypes are all complementary approaches for designing and evaluating human-centered technological system~\cite{Hutchinson:techprobe:2003,Sanders:Probes:2014}. A framework by Sanders~\textit{et al.}~\cite{Sanders:Probes:2014} emphasizes the particular importance of probes for conducting pre-design work that aims to find \textit{``inspiration in users’ reactions to [the probe's] suggestions''.} They contrast probes to  prototypes that aim to test a specific idea and tend to be used in evaluative research~\cite{Lam:Scenarios:2012}.} 
Prior research has demonstrated that design probes can be especially useful for creating a human-centered experience for machine learning~\cite{Broadley:VisualisingHD:2013,Stumpf:InteractingMW:2009}, which becomes increasingly essential in light of the increasing diversity of those using (and impacted by) AI/ML technology. While methods for developing and using design probes vary, a recent framework by  Subramonyam~\textit{et al.}~\cite{Subramonyam:Prtotyping:2021} has synthesized these common approaches into a Model Informed Prototyping Workflow that \textit{``combines model exploration with UI prototyping tasks''}. Recently, they also propose the use of \textit{data probes} to co-create Ml/AI interactive experiences for non-expert analysts~\cite{Subramonyam:cocreationg:2021}. The mechanisms for capturing user interactions and an ML model's response varies by according to the approach. For example, design probes can be coupled with  Wizard of Oz approach that simulates ML model behavior~\cite{browne:ml_wizard_of_oz:2019,Maulsby:Woo:1993,Dove:MLWoo"2017}. Prototypes like Gamut~\cite{hohman:gamut:2019} automatically capture interactions and display model responses.
\rev{While visualization researchers use a variety of approaches, we note a preference for high-fidelity prototypes assessed in summative evaluations~\cite{Lam:Scenarios:2012}. We see opportunities to further explore approaches for co-creating visual and interactive ML systems together with users in pre-design research~\cite{Brehmer:design:2014}.}

%% file: tex/elicitation_study.tex
\section{Data and Visual Design Probes}\label{sec:design_probes}

In this section, we describe the construction of a data probe and a set of visual design probes that we use in a subsequent elicitation study.
Supplemental materials for our probes have been made available online: \url{https://osf.io/8wbgf}.

\rev{\textbf{The primary objective of our data and design probes is to elicit interactions with a machine learning model with the intent of steering the model's training.} \revMR{We do not limit ourselves to uncovering only novel interactions, but also to collect the broad scope of users' desired interactions.} In addition to collecting these interactions, we wanted to also explore the factors motivating those interactions including reasons why interaction is not desirable. \revMR{Collecting both desired interactions and users' expressed intents allows examine and contrast what is both desirable and feasible.}}

\rev{The possible design space for these data and visual probes is vast. We prioritize understanding how users directly manipulate data-encoding marks of visual encodings to steer an ML model.  The result is that our visual design probes emphasize different chart types, similar in spirit to the approach taken by Saket~\textit{et. al.}~\cite{Saket:2020:Direct_Manipulation}, and not all aspects of the interface. Instead, we use our probes to elicit the other aspects of the user interface (i.e., performance diagnostic plots, explainability features, etc.) they believe would be useful. Finally, we aimed to minimize potential  attribution issues that impact visual  IML systems~\cite{Boukhelifa:eval_attribution:2020}.}~\rev{We created our data and visual design probes with these objectives in mind.}

\subsection{Data Probe}\label{dataprobe} 
\rev{Data probes are a useful tool for encouraging divergent thinking about the behaviors on IML systems, considering their boundaries and limitations, and imagining (or proposing) alternative behaviors~\cite{Subramonyam:cocreationg:2021}. For our purposes, data probes are useful for reflecting on the efficacy of the different visual encoding choices and the interactions they afford. We created a design probe for a synthetic movies dataset. We use synthetic data, over an actual dataset, because it allows us to control the shape and distribution of the data, including the introduction of noise and errors, to probe end-users perceptions and their desired interactions.}

\vspace{1mm}
\noindent{\textbf{Overview.}} \rev{Our data probe explores the output of a topic classification model.} We created a dataset of 50 movies that we compiled from IMBD and Rotten Tomatoes, harvesting their title, a brief synopsis, and primary genre (i.e., its classification label). We select five movies belonging primarily to one of ten genres:  Action, Animation, Comedy, Drama, Fantasy, Holiday, Horror, Romance, Sci-Fi, and Thriller. 
\rev{To construct our synthetic dataset, we first deliberately introduce errors into the labels. For example, \textit{Dr. Strangelove} is a comedy movie, but we set its predicted genre as a romance; \textit{Die Hard} is a `Holiday' movie but is predicted to be `Action')}\footnote{The movie `Die Hard' takes place on Christmas Eve and has prompted a (sometimes facetious, sometimes serious) debate of whether it is a holiday film or not. Among our participants, 60\% said it was solely an action movie, 35\% thought it was both an action and holiday movie, and 5\% were unsure.}.~\rev{Next, we use a Gaussian Mixture Model (GMM)s to simulate the classification probabilities as well as a set of two-dimensional XY coordinates for each data point; the 2D coordinates are a substitute for computing a dimensional reduced representation of the dataset. We explored different parameter settings for the GMM and its components, to modify the number spread of each cluster, their proximity and orientation to each other, and the extent that they overlap (see supplemental materials). We selected one set of simulated results for the final data probe.}


\vspace{1mm}
\noindent\textbf{Design Considerations.} While movies can belong to multiple genres, we deliberately imposed a single classification label per movie to observe if participants would be motivated to modify this. If participants were to introduce multiple labels per movie, then they would be introducing a change to the model that would go beyond simple parameter updating. We similarly use errors and ambiguities as prompts for discussion around the data and the model. In both instances, we are interested in assessing if participants would discover these prompts via the visual design probes and how they would respond to them.

\rev{The overall size of the data probe is smaller than a typical ML workflow, which features thousands, millions, or even billions of points depending on the dataset. \revMI{The smaller size of 50 points was pragmatically selected to make the design and interaction of the visual design probes feasible, whilst still having diversity and variation in the dataset. Visualization research has not proposed a singular dataset for clustering data, nor has prior research concerning clustering with visualization discussed the relationship between dataset size, total marks displayed, and interactions. Lacking such guidance, we opted for a dataset size amenable to our research objectives}. However, we argue that while datasets are massive, end-users rarely interact with all data points at once. They may filter or aggregate information, leaving them with a more reasonable number of points to interact with.  While the design of our data probe may not be suitable for answering all possible questions about model steering through visual encodings, it does allow us to explore the fundamental questions of how end-users interact with data encoding marks where each mark represents one data point.}

\vspace{1mm}
\noindent\revMR{\textbf{Generality and Transferability.} We derived the probe data from two commonly used machine learning datasets with modifications to deliberately introduce errors and ambiguity. \revMI{Moreover, the movies that we selected for our probe are popular such that the majority, if not all, participants were familiar with them and their premise. The choice of dataset also served to ameliorate two potential confounding factors in this study: dataset characteristics and participant's dataset familiarity.}  Our dataset is not intended to serve as such a benchmark, others may find our approach transferable to other datasets and applications.}



\subsection{Visual Design Probe} 
\rev{We used the data probe to create a set of visual design probes. Visual and interactive ML interfaces can be complex and multifaceted and include modalities for the user both to interact with data, through both existing and bespoke visual encodings~\cite{Yuan:survey:2021}, and get feedback on the effects of their interactions~\cite{dudley:UI_IML:2018}. The complexity of these interfaces can also lead to attribution challenges~\cite{Boukhelifa:eval_attribution:2020}; it is not clear what the value of individual interface components is. We construct visual design probes that examine the relationship between the visual encoding design choices and participants' interactions with individual data marks. Our visual design probes would constitute \textit{one component} of an overall IML system and we use these probes to also elicit from users \textit{what the other interface components should be}.} 

\begin{figure}[h!]
    \centering
 \includegraphics[width=\linewidth]{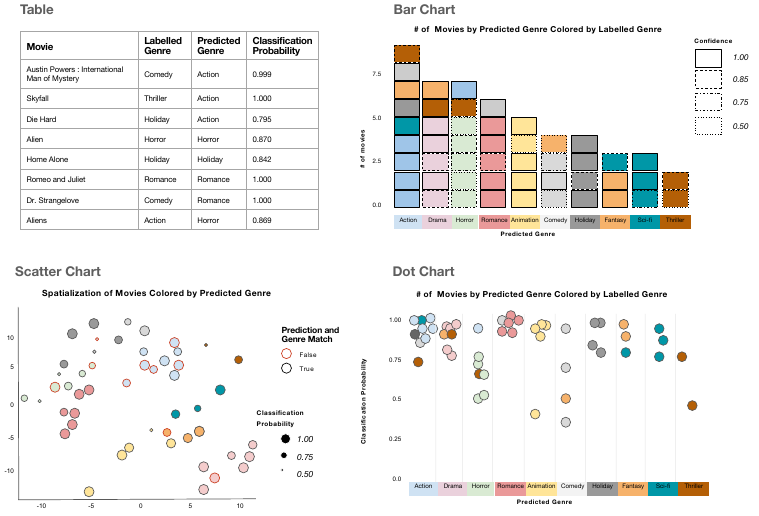}
    \vspace{-5mm}
    \caption{Visual design probes created for our elicitation study}
    \vspace{-3mm}
    \label{fig:encoding_charts}
\end{figure}

\vspace{1mm}
\noindent{\textbf{Overview.}} We create visual design probes for a simple text classification problem. We mapped the elements of the data probe to marks and channels to a set of four visual encodings: a table, a bar chart, a scatter chart, and a dot chart.  While the table is potentially an unusual type of visual encoding, it remains a common way to view and report on machine learning results and serves as an important baseline. Encoding choices are summarized in~\autoref{tab:encoding_table} and the four visual design probes are shown in~\autoref{fig:encoding_charts}.

\begin{table}[t!]
    \centering
    \includegraphics[width=\linewidth]{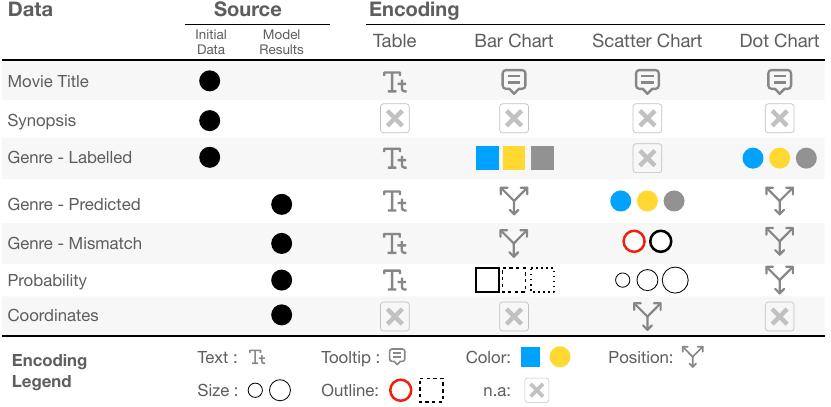}
    \vspace{-2mm}
    \caption{Mapping of elements of the data probe to visual encoding}
    \label{tab:encoding_table}
\end{table}

\vspace{1mm}
\noindent{\textit{Table:}} Each row displays one movie along with its labeled and prediction genres and the classification probability.

\vspace{0.5mm}
\noindent{\textit{Bar Chart:}} Each rectangular mark in the bar chart represents a single movie. The marks are vertically stacked and positioned along the x-axis according to the predicted genre. The x-axis is also sorted by the number of movies predicted to belong to the genre. The color of the individual marks represents the labeled genre. The model's prediction probability is shown via the mark's outline; a solid outline indicates higher confidence, whereas the dashes indicate lower confidence.

\vspace{0.5mm}
\noindent{\textit{Scatter Chart:}} Each point in the scatter chart is a single movie. The marks are positioned according to x and y coordinates that are simulated from the GMM. The marks are colored according to their predicted genre and sized according to the model's predicted probability; smaller mark sizes indicate lower prediction probabilities. A red stroke outline around the marks indicates whether the predicted genre matches the labeled genre (black outline) or not (red outline). The scatter chart is also intended to represent lower dimensionality data representations or embeddings that are routinely presented alongside ML models.

\vspace{0.5mm}
\noindent{\textit{Dot Chart:}} Each point in the dot chart is a single movie. The points are positioned along the x-axis according to their predicted genre and along the y-axis according to the predicted probability. As with the bar chart, the genres are sorted according to class size from largest to smallest. The colors of the point marks encode the labeled genre.  

\vspace{0.5mm}
\noindent{\textit{Alternative Chart Types:}} We recognize that different individuals may expect different visual encoding types, or may not find the default encoding choices effective. 
In considering how we implement our probes we deliberately sought to provide participants with the tooling to challenge our design choices and even propose new visual encodings.  

\vspace{5mm}
\noindent{\textbf{Design Considerations.}}  We choose these three visual encodings (and table) because, based on our experience, they are common ways to visual ML model data and results. Others have also previously explored these visualizations in prior work of direct manipulations of data encoding marks~\cite{Saket:2020:Direct_Manipulation}. The goal of consistently encoding data across probes is to reduce the potential for anyone visualization to have an outsized effect on the interactions. Others who leverage data and visual probes in their work may find greater variability in the probe design desirable. However, in a pilot study (Section~\ref{sec:pilot_study}) we found that too much variability was disorienting. We modified our probes ahead of the full study (Section~\ref{sec:full_study}) to the versions shown in~\autoref{fig:encoding_charts}.

\revMI{We use the same visual design probes for both ML and non-ML Experts. While these two groups have different expertise, which influences their needs, we believe that visualizations are a common and reasonable baseline for both groups. We rely on both the expressivity of our probe medium and our elicitation study structure to capture additional elements participants would need, as well as, the similarity and differences between these groups. }
\vspace{-2mm}
\subsection{Visual Design Probe Implementation}\label{probe_imp}
\rev{We developed visual design probes that are situated between a paper prototype and a web-based app. Through our prototype, an end-user can interact by directly manipulating marks in the chart (i.e., change its position, color, size, etc.). The responses to these interactions are driven by a human 'wizard' (i.e., the study administrators) and assessed by the end-users.}

\vspace{1mm}
\noindent{\textbf{Overview.}} \rev{We considered different materials for implementing our visual design probes and found Google Slides to be a surprisingly compelling medium to meet our goals. As a medium, Slides allowed us to implement the different visual encodings, provided a way for participants to easily change them (i.e. change position, color, size, shape, etc.), and introduce new encoding elements. It also allowed us to securely share and save individual sessions with different users. The slides environment was also familiar, reducing concerns about the effects of systems novelty~\cite{Boukhelifa:eval_attribution:2020}. Finally, Slides was also collaborative allowing for an administrator to create interactive responses to end-user interactions.} 

\vspace{1mm}
\noindent{\textbf{\revMR{Affordances of Visual Design Probes.}}}
The four visual encodings in~\autoref{fig:encoding_charts} were implemented in Google Slides with individual marks and channels capable of being directly manipulated (i.e., moving a point, changing its size, or color). \revMR{We chose this medium because it provided considerable encoding and interaction affordances. At a minimum,} we were able to support the full spectrum of interactive operations reported in Saket~\textit{et al.}~\cite{Saket:2020:Direct_Manipulation} (i.e., modifying position, size, color, height, and width encodings) in addition to others such as adding text annotations, new shapes, or comments. We also created tooltips for each mark by making use of hyperlink previews. 

\revMR{The goal of our visual design probes was to hone in on direct manipulations with data encoding marks, not a whole system. 
We believe this focused approach can collect useful empirical data by providing a canvas for expressing desired interactions with visual encodings. Moreover, the affordances of our probe can provide sufficiently generative power to meaningfully extract the relationships between data, interactions, and desired model updates. } 

\vspace{1mm}
\noindent{\textbf{Simulating Model Modifications.}}~\rev{We used a Wizard-of-Oz~\cite{Maulsby:Woo:1993} approach to simulate model updates and system behaviors.} \revMR{Our Wizard made two assumptions about participant interactions with visual design probes. First, any interactions with data encoding marks were valid. For example, if a user wanted to change the size, position, and/or color of a mark to express the same intent, such as changing a class label of one or more data points, these were all valid actions.  Second, participants could introduce \textit{new} data marks and the wizard could learn from them. For example, adding an ellipse around a group of points to denote a class.  Given these interaction assumptions, the possible responses by the Wizard were broad, but allowed us the flexibility to collect a variety of participant perspectives as part of a pre-design empirical phase. Notably, we made no assumptions towards the back-end ML model for the Wizard's responses. 
We assumed that some of the participants' interactions could conform to existing models, including the GMM we used to generate the data, but we also wanted to capture possible instances were they did not. } 

\rev{A study administrator would craft a model's possible responses in response to participant interactions and the visual encoding choice. For example, if an end-user changed the position of the mark, a study administrator could respond by changing its color or moving additional points that were proximal to the one that moved. \revMR{The administrator prompted participants for their expected rate of update, for example, with each interaction or following a sequence of interactions. Lastly, before the model was updated participants were asked to indicate their expected response. The wizard used these utterances to craft a response by changing marks in the visual encoding. \revMI{We did not have a prescribed procedure for how the Wizard should interpret the utterances; it was left to them to interpret based on session and the context.}~Participants were not informed that their utterances were used to carry out the response. After the update was carried out, participants were prompted again to see if the changes aligned with their expectation. Given the medium, participants were aware that a human was driving the updates, but they did not know how the updates were being made.} This approach allowed for co-creation processes of ideating around different individual expectations of model responses.}

\section{Elicitation Study}\label{sec:elicit_study_overview}
We conducted an elicitation study using the visual design probes. Elicitation interviews are a qualitative technique that uses an iterative approach to reveal increasingly granular details of a participant's experience within a situated scenario, in this case steering a classification model~\cite{hogan:2016:elicitation}. \revMR{Our goal of the elicitation study is to explore ``approaches, processes, and subjective experiences''~\cite{hogan:2016:elicitation} of interacting with ML models. We argue that this approach allows us to explore the \textit{semantic distance}~\cite{Hutchins1985DirectMI} between a variety of encoding and interaction affordances and the user's intended model modifications. To that end, we seek to not only elicit novel interactions, but to examine \textit{how} interactions are linked to possible modifications of ML models and \textit{what} the intended effects of these interactions are. }  

\vspace{-3mm}
\subsection{Research Questions}\label{subsec:research_Q}
We are motivated by the challenges that domain experts experience when needed to update machine learning models, often without the use of code~\cite{Crisan:2021:automl,Xin2021WhitherAU,Qian:2018:MTeaching}. We also wanted to understand if and how the needs of these domain experts differ from individuals that have formal ML training. Our study sought to examine the following research questions : 

\begin{itemize}
    \setlength{\itemsep}{0pt}
    \item \textbf{RQ1} : \textit{How} do participants choose to interact with marks of visual encodings to steer a model?
    \item \textbf{RQ2} : \textit{What} are participants' perspectives or concerns towards these interactions?
    \item \textbf{RQ3} : \textit{What differences} exist (if any) between ML and non-ML experts? 
\end{itemize}

\revMR{We use a the same visual design probes to assess these research questions with both ML and non-ML experts. While the needs of both groups are different at face value, we were interested in understanding to what extent and in what ways expectations of these two groups overlap or differ relative to a common baseline. 
We posit that using a common baseline as part of pre-design empirical work~\cite{Brehmer:design:2014,Sanders:Probes:2014} helps to calibrate and evaluate IML interfaces. }
\vspace{-2mm}

\subsection {Participant Recruitment and Background}\label{subsec:recruitment}

\noindent \textbf{ML and non-ML expert definition.} We use the definition of a non-ML expert articulated by Qian~\textit{et. al.}~\cite{Qian:2018:MTeaching}, which is that they are \textit{``people who are not formally trained in ML and are actively building ML solutions to serve their needs in the real world''}. This definition excludes lay people that, in addition to lacking forward knowledge, do not actively build or develop ML workflows even though they may encounter ML models in their daily work~\cite{Swati:design_nonexperts:2021}. We define an ML expert to be someone with formal training (i.e., undergraduate or graduate-level education) in ML or DS.

\vspace{1mm}
\noindent\textbf{Recruitment.}~\rev{We recruited an even number of participants with and without ML expertise to participate in our study. We posted calls to participate on Twitter and various machine learning and data science Slack Channels, in addition to using snowball sampling~\cite{Creswell_Poth_2018} to recruit and screen eligible participants and determine if they were classified as ML or non-ML experts.}
We also sought to have some gender balance in our recruiting, specifically reaching out to individuals that identified as female to participate. We recruited a total of 20 participants; \rev{ half of the participants identified as female or non-binary}.  The initial four recruits were used in a pilot study (Section \ref{sec:pilot_study}) that served to both refine the data and design probes and also to establish an appropriate sample size. 
Following the initial full study data analysis, we reflected on the findings and established that we had reached analytic saturation. Based on this analysis we determined another round of recruitment was not required. 

\vspace{1mm}
\noindent{\textbf{Background.}} A total of 8 of 20 (40\%) identified as data scientists, 6 (30\%) as data or machine learning engineers, 6 (30\%) as software engineers,  3 analysts (15\%),  and finally 3 (15\%) as visualization researchers. A total of 6\% of individuals did not identify with any of these roles. Note that participants could select more than one category to report their role. 
Aligning with our definition of ML and non-ML experts, all participants had regular exposure to ML models but reported a varied focus on model work in their daily activities. In total 8 (40\%) reported that training machine learning models were a daily component of their job; note that building an ML solution does not require training a model, as it is possible to use pre-trained models. An additional 10 (50\%) report primarily interpreting machine learning results in their daily work. 


\vspace{-2mm}
\subsection{Session Procedures}\label{subsec:study_procedure}
\rev{\noindent\textbf{On boarding.} \revMR{All sessions took place over a video conferencing platform.} We participants were provided with an overview of the study after which we obtained consent to participate to record audio and video of the session. Participants were then provided with a unique study link to access a set of training materials and the study probes. For the training materials, participants were given an overview of the data probe (Section~\ref{dataprobe}). They were also provided with a simple bar chart encoding and asked to complete a set of basic interaction tasks (i.e., move a point, change its color or position, hover to reveal a tooltip) to familiarize themselves with the affordances of Google Slides. As most participants had experience using Slides, or some other presentation tool, all found this task to be straightforward.} 

\vspace{1mm}
\noindent\textbf{Interacting with Visual Design Probes.} \rev{Participants were then presented with the design probes(\autoref{fig:encoding_charts}) in one of four possible orders (Section~\ref{sec:pilot_study}). For each visual encoding, participants were given a brief overview of the data that is encoded and were provided with a set of possible interactions (i.e., move a point, change its color).} \textit{Importantly, participants were informed that it was valid to indicate that they did wish to make any modifications.} Participants were prompted to speak out loud to describe what they were doing and why. \rev{In response to both participant interactions and responses, the study administrator would simulate a response (see Section~\ref{probe_imp}). These were sometimes simple responses (filtering, sorting/re-ordering, or highlighting) or were model responses (updating the positions of other points, reclassifying data according to new labels, updating classification probabilities).  Lastly, participants were encouraged to consider what additional feedback would help them understand and contextualize their actions; if they chose not to interact, they were asked to discuss why}. 

\vspace{1mm}
\noindent\textbf{\revMR{Facilitating Model Updates.} }\revMR{Model updates proceeded at a cadence in accordance with participants' stated preferences (see Section~\ref{probe_imp}); in practice, nearly all participants wanted updates after completing a sequence of interactions, not with each individual interaction. For the update step, the wizard would manipulate individual data encoding marks, or groups of marks, depending upon the participants' interactions and their expressed model steering intents. Given the size of the data probe and the simplicity of the visual design probes the `model update' could be carried out in under a minute in most instances. Participants could see individual marks being moved in real-time. At the end of the update, indicated by a prolonged pause, participants were asked to comment on the `responses' and whether the cadence of change and magnitude of change aligned with their expectations. Once again, the specific responses were tailored to each individual and the particulars of that session. The process of model refinement would continue until participants indicated they wished to conclude.  }  

\vspace{1mm}
\noindent\textbf{Wrapping Up}. Participants were asked to reflect on their experience overall and provide any final thoughts.

\subsection {Data Collection and Analysis}\label{subsec:data-collection}
We collected and analyzed approximately 20 hours of recordings. We retained modifications participants made to the design probes and took notes throughout the sessions. Video recordings captured participants' interactions with the design probes and audio was transcribed for analysis. We manually annotated the video recordings for participants' direct manipulation of visual encoding marks. We record single instances of an interaction per visualization, not per mark; for example, if a participant changes the color of 10 different point marks in the scatter chart, this is tallied as just one instance of semantic interaction and not 10. We use descriptive statistics to summarize the types and frequencies of interactions across sessions. We use participant utterances to add context to both the purpose of the interactions, and any hesitations.

\subsection{Session Overview}
\revMI{Sessions lasted on average 50 minutes; the shortest session was 40 minutes and the longest was 1h and 7 minutes. Participants spent on average 5 minutes and 32 seconds with each encoding; administrators checked in with participants either at the five-minute mark or when they observed a prolong paused (one minute or greater). In total participants spent 22 minutes and 40 seconds (roughly 50\%) of the overall session time across all four encodings. The rest of the session duration consisted of consent, overview, practice, and wrapping up and addressing technical issues (video, connectivity, screen sharing) if any arose.  On average participants spent 5m33s with the table, 5m31s with the bar chart, 
 5m56s with the scatter chart, and 5m17s with the dot chart. The interactions with the scatter chart were slightly slower because it took more time for participants to identify a mark they were interested in with the hover action. The longest a participant spent with any one encoding was approximately 9 minutes. We did not precisely time the responses of the Wizard, but did not note any response that took especially long to craft for any encoding. This is largely because the Wizard was reasonably familiar with correspondence between marks and certain movies and given the dataset size was able to quickly respond.}

\revMI{Participants did not continuously interact with an encoding during a session, instead they would perform as a set of interactions after which they would be prompted by the study administrators to explain the meaning behind them. Participants varied in how much or little they chose to discuss their interactions; again while we lack precise timing, anecdotally some participants were more verbose than others, which may have contributed to longer time spent with an encoding.}


\begin{figure*}[ht!]
    \centering
    \includegraphics[width=0.95\textwidth]{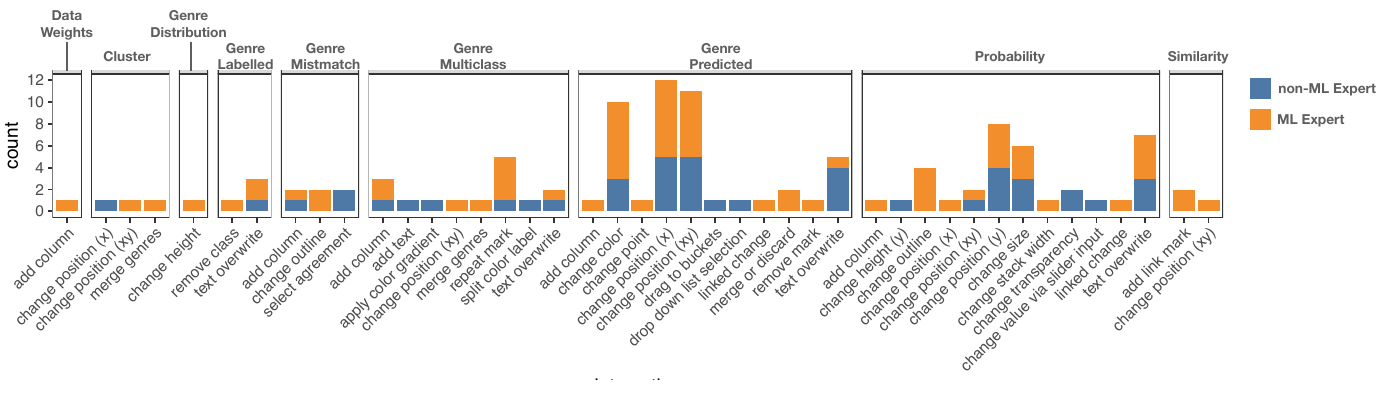}
    \vspace{-7mm}
    \caption{Frequency of  Interactions (x-axis) and their Data Targets (facets) elicited from a study using the visual design probes}
    \vspace{-3mm}
    \label{fig:res_details}
\end{figure*}

\subsection{Pilot Study} \label{sec:pilot_study}
We conducted a preliminary pilot study with four participants to refine the visual design probes (i.e., spelling errors, complex flow, and design inconsistencies).  The pilot study also revealed the ordering effect of our visual probes, which introduced an anchoring effect~\cite{valdez:anchoring:2018}.
To mitigate this effect, we grouped the table (T), bar chart (B), scatter chart (S), and dot chart (D) into two sets and varied the order in which they would be presented to the participants as follows: [(TS)(DB)], [(TD)(BS)], [(ST)(DB)], [(DB)(ST)].  Given these presentation orders, our study would need to recruit at least eight participants (4 ML and 4 non-ML). We wanted at least each order to be viewed twice within each group, which resulted in a minimum study size of 16 participants.  

%% file: tex/full_study.tex
\section{Elicitation Study Findings}\label{sec:full_study}
We summarize the results of our elicitation study according to our research questions~(Section~\ref{subsec:research_Q}).

\begin{figure*}[t!]
    \includegraphics[width=\textwidth]{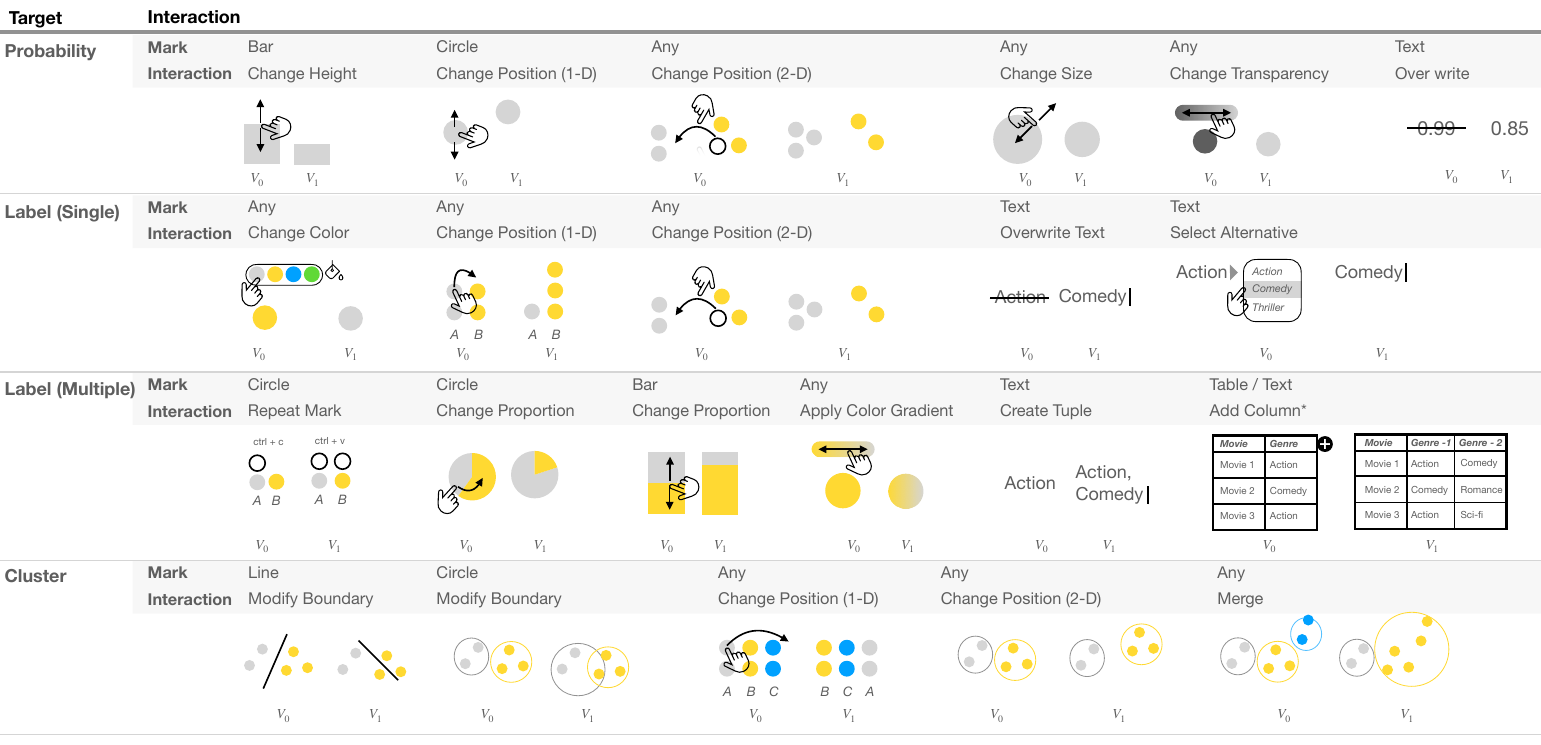}
    \caption{A subset of common target-interaction pairs elicited from participants in our elicitation study. The difference between the initial ($V_0$) and resultant visualization ($V_1$) can be used to refine an ML model without directly exposing the person to its parameters.}
   \label{fig:results_main}
   \vspace{-3mm}
\end{figure*}

\begin{figure}[t!]
    \centering
    \includegraphics[width=\linewidth]{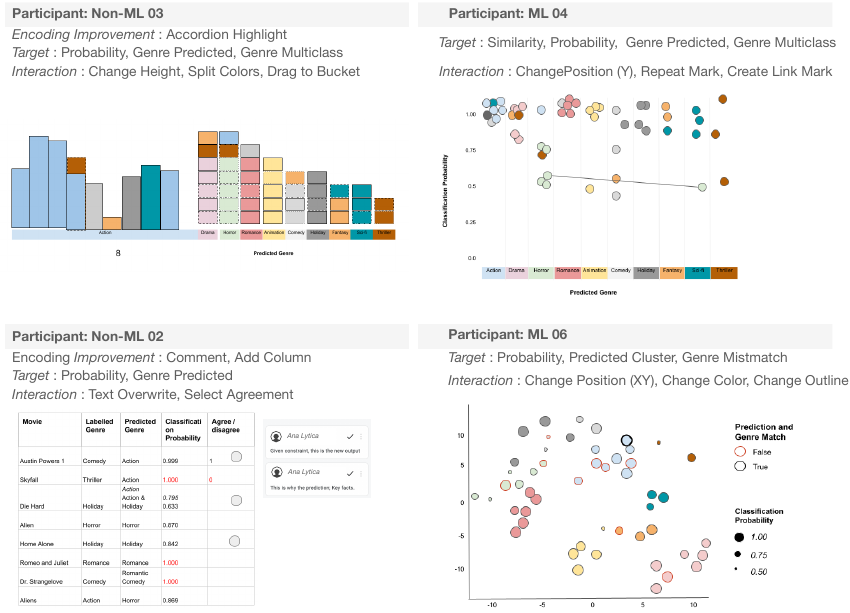}
    \caption{Examples of larger changes participants made, which include changing the visual encoding, adding new elements, and adding new components to the visual interface like comments}
    \label{fig:ux-changes}
    \vspace{-5mm}
\end{figure}

\subsection{RQ1:  A Spectrum of Interaction}
We summarize participants' interactions with visual encoding marks as target-interaction pairs. \textit{\textbf{Targets}} are the data attributes that participants sought to modify and \textit{\textbf{interactions}} are the mechanism they used to do so.  Participants demonstrated a variety of strategies and consistency when manipulating marks of our visual design probes. The full set is enumerated in~\autoref{fig:res_details} and a subset of the most common target and interaction pairs are shown in ~\autoref{fig:results_main}. Across all of our study sessions, we recorded a total of 114 examples of interactions that we summarize as a set of 43 unique target-interaction pairs.  These interactions were not gathered uniformly across participants; two (one ML and one non-ML) provided no interactions at all (per the session procedures (Section~\ref{subsec:study_procedure}) it was valid to indicate no desire to interact). Conversely, one participant demonstrated 16 (of 43) unique interactions; the median was 8.  Participants targeted their interactions both at individual data encoding marks, and sets of marks. 
Critically, we found the interactions were deliberate and that participants could articulate their interaction intents and expectations.

\subsubsection{Data Targets}\label{subsec:data_targets}
We identified a total of 9 data targets; this number exceeds the total number of raw attributes in the dataset because participants also introduced new attributes through the encodings. As an example, one participant added a column to the table probe to introduce a weight to each movie in the dataset. The participant wanted the weight column to inform the model of data points to prioritize in retraining: \textit{``I want to tell the model to worry about these points and don’t worry as much about getting these [others ones] right [...] which would allow me to say, I really care about these and I don’t so much care about those''} [ML:07]. 
One participant indicated wanted to augment the data through the interactions, a reflection of the expressivity of our design probes: \textit{``when I think about what I am feeding the machine learning model there, it's essentially creating another dimension for it to run through[..]  I am saying `Here’s a user dimension that I want you to consider [..] with the other dimensions that you have'}[ML:04]. Lastly, nearly all participants also introduced multiclass labels for each movie, which for analysis we treated as separate from the original single class label (the primary genre). 

Considering both initial datasets and additional participant-derived attributes collectively, we found that the most common targets of interactions are primary and predicted labels (genres) and model probabilities (\autoref{fig:res_details}). However, we also observed instances when participants also wanted to indicate when they agreed with the predicted label instead of the initial label: \textit{``I also have this urge to say that I like how this is being classified [over the primary]''} [non-ML:06]. These findings show that participants are not just seeking to update the ML model parameters. They are operating on a meta-modeling level by saying that the way the data is modeled might not be true to the meaning of the data itself; they are challenging the ``ground truth'' of the given data. 
Moreover, although our data probe is constructed for a simple classification model, our findings show that the scope of modifications participants wish to make would push the boundaries of such a  model by challenging the data assumptions it was trained on. 


\subsubsection{Mechanisms of Interaction}\label{subsec:interactions}
Participants demonstrated a variety of strategies for interacting with data targets (\autoref{fig:res_details}). The most common types of interactions were changing the position, size (or height), and color of the mark. These types of interactions were common to the bar, scatter, and dot plots. With the table data, participants would frequently overwrite existing text, for example, to modify a movie's genre or its prediction probability. Interestingly, participants also introduced graphical elements to the table. The most common was coloring the rows to reflect the classification probability. Some participants added marks to new columns (see~\autoref{fig:ux-changes}) that they wanted to use to modify the data weight of the row. 

Participants explored encoding designs beyond these defaults, a set of examples from four participants is shown in~\autoref{fig:ux-changes}. One participant [non-ML:03] used the design prob to prototype a ``accordion view'' with the bar chart; surprisingly, and independently, another participant [non-ML:05] had a similar request but described it instead of implementing it. Participant [non-ML:03] proposed that clicking on a column within the stacked bar chart would trigger the individual marks to reposition themselves next to each other. The participant demonstrated how they would then inspect the individual marks and modify their heights (prediction probability) and even introduce new genres by recoloring segments of the mark. This was the most complex strategy we observed \revMR{and indicated to us that the visual design probes provided affordances for participants to creatively express themselves. Such expressivity is useful for challenging existing visual design ideas, as was the case for our probes, and iterating on encoding and interaction design}. Other new marks participants proposed included adding links between marks (ML-04) to indicate similarity, and adding columns to the tables to add more data, such as data weights (see Section~\ref{subsec:data_targets}). Participants introduced new marks by repeating an existing one. For example, when wanting to express that a movie belongs to multiple classes, several participants choose the strategy of copying a mark and pasting it into multiple genres. Lastly, participants also explored introducing new, or alternative, channel encodings such as color gradients and transparency to encode multiple classes and prediction probability, respectively. 
Overall, both ML and non-ML experts used similar types of interactions and targeted similar data attributes for model steering.

\vspace{1mm}
\noindent\textbf{Reasons not to Interact.}\label{no-interaction} Only two (of 20) participants declined to interact with any of the visual design probes. One of these participants was a non-ML expert and expressed concerns about injecting bias (other participants also expressed these concerns and we elaborate further in Section~\ref{bias-concerns}). The other participant was an ML expert and felt that visualizations were primarily for communication and not for model steering - they preferred to code a solution, but this is not an option for many non-ML experts. Other ML expert participants saw the value of interacting through visualizations to explore model steering for certain tasks. These ML participants also indicated instances when they did not want to refine the model but completely retrain it or select a new model. Three participants indicated that they would want to retrain the model on only the misclassified results. Two said that they would use the visualizations to decide whether to choose a different model. In these instances, ML experts felt it was better to drop into code instead of interacting through an interface.

\vspace{1mm}
\noindent\textbf{Additional Properties of Interactions.} We observed that individual interactions could also be both compounded and hierarchical. Compounded actions affect multiple data targets simultaneously, for example, moving a mark from one position to another, with the intent to change its class, should be accompanied by an automatic change to its probability (higher for the new class, lower for the new class), and vice versa. Actions were hierarchical when the participants wanted to make changes that should cascade down to individual points. Two participants described hierarchy actions via anchor points; these were points participants chose to represent a cluster and any modifications to those anchor points should be applied to others near it.


\subsubsection{\revMR{Feasibility and Utility of Interactions}}
\revMR{We identified two factors impacting the feasibility of implementing the interactions participants demonstrated via our probes: whether it was reasonable to implement the interaction and whether it was possible to use the data generated by the interaction to update an ML model. First, some interactions, although possible to implement, might not be reasonable to scale. An example is using gradients to indicate multiple classes (i.e, a movie belongs to two categories). The design space we collect offers multiple alternative strategies for encoding multiple classes, elicited by participants, that may be more favorable according to multiple participants and reasonable to implement, such as repeating a mark (see~\autoref{fig:res_details}). Another factor of feasibility is whether the data generated by the interaction could actually be used in an ML training regimen.} \rev{ For example, participants expressed introducing new class labels, having multi-class assignments, as well as introducing hierarchy to data points. Not all classification models would be able to accommodate such interactions and prototype systems that failed to anticipate them would not be able to support them at all.} \revMR{In this case, the probes can suggest further directions for expanding ML models to incorporate user feedback. Importantly, it may also be used to help refine an ML model backed in a visual IML system. For example, if participants consistently introduce multiple classes or data hierarchy, as participants did in our study, then we could choose a model that supports the incorporation of such data. It may be possible, but challenging, to revise a more complex prototype to align with user expectations.}

\revMR{We can also consider the utility of interactions surfaced.} \rev{We observed that the interactions surfaced by our visual design probes have also been identified in prior studies, which lends validity to the types of interactions our probes are able to elicit.}  For example, others have reported~\cite{Drucker:iCluster:2011}, we also observed behaviors of grouping points and annotating clusters, in one instance to create completely new clusters with labels that did not exist in the initial (training) dataset. When we compare our results to the comprehensive summary of cluster interactions in Bae~\textit{et al.}~\cite{Bae:iClustering:2020}, we also find overlapping examples of participants adding or merging, modifications to the cluster's structure (i.e., moving marks in and out of clusters), adding constraints by anchoring marks to clusters, as well as correcting and validating perceived errors (i.e., Dr. Strangelove is not a Romance film, but a comedy). Actions to overwrite text in tables are similar to the behaviors of systems like StarSPIRE~\cite{Bradel:Starspire:1995}. Moreover, some interactions that users undertook evoked model steering mechanisms that have also been proposed by prior works. Participants' desire to re-weight points is an updating approach that appears to be similar to iteratively refining a weak learner via boosting~\cite{Schapire:boosting:2005}, but with manual intervention from a human. Re-weighting data in accordance with user interactions has also been used in other IML systems~\cite{Endert:2012:SemanticOG,el-assady_progressive_2018}.

\revMR{While there are limitations on the full spectrum of interactions our visual design probes could capture, overall, the alignment between our findings and prior work is an encouraging demonstration of the probes' capacity for capturing useful interactions. Moreover, we can identify novel interactions with respect to the combination of targets and actions that prior work has not identified. For example, prior work does not identify probability as an interaction target for clustering and classification, yet the use of visual design probes surfaced it and identified several consistent interaction strategies (\autoref{fig:res_details} and \autoref{fig:results_main}).} 

\subsection{RQ2: Participants Perceptions and Concerns}
In addition to capturing participants' interactions, we also dive more deeply into the factors that influence their interaction choices, including their hesitations.~\rev{What is noteworthy about participants' responses is the sensitivity of design choices on participants' willingness to interact with a system. For example, building a system with diagnostic plots to contextualize the effects of interactions would have been not useful to others, especially non-ML experts. Prior work has also shown that the magnitude of model changes is not always easy to capture~\cite{Crisan:2021:text}. Overall, these findings show the value of design probes to anticipate challenges proactively and influence the final system design.}

\subsubsection{Factors Impacting Interaction}
\noindent\textbf{Effects of Encoding Design on Interactions}\label{subsec:encoding_effects}
\rev{The choice of visual encoding impacted how `inviting' it was to engage in an interaction.} For one participant, the table was more inviting  because when they see visualizations they are \textit{``not sure how moving things around [in the scatter plot] change anything [and I]  see a table as a natural place to provide row-level feedback.''}[non-ML:06]. For other chart types, certain interactions were less desirable. For example, some participants would repeat a mark to indicate a movie belonged to two or more genres. Most participants performed this mark repetition interaction with the dot and scatter chart encodings but did not do so with the bar chart even though it was possible. When prompted, one participant offered that \textit{``it just feels wrong''}~(ML:04). There was also a general consensus among both ML and non-ML experts that the scatter chart was the most difficult to reason with. For many participants, the axes of scatter charts are tied to single concrete attributes (i.e., age, height, eye color, etc.) and it was difficult to reason about axes that represented embeddings or dimensionally reduced data. Participant non-ML:05 described the dissonance as `L1-interference', not to be confused with L1-regularization, which occurs when concepts are translated from one's native language (L1) to a new language (L2). Their rationale was that using visualizations to steering models was like learning a new language that was similar, but different, from one that they were accustomed to.

\vspace{1mm}
\noindent\textbf{Anticipating the Effects of Interactions.}\label{interaction-hesitancy} The choice of visual encoding was also tied to participants' ability to understand and contextualize the effects of their interactions. Participants expressed preferring simpler visual encodings because it was easier to anticipate the effects of their actions:\textit{`` I feel like engaging with the other charts I feel more confident [..] because[..] when I change position [on the scatter chart] I will change the distance to all the other points.''} [ML:06].
~ Visual encodings also evoked `degrees of freedom' that a participant could use to interact, with too many being overwhelming: \textit{``I think I prefer this vis [bar chart]. There’s just too much freedom in the scatter plot''}[ML:05].

\rev{One way to ameliorate this challenge is to include additional visuals for contextual information. Our visual design probes were intended to give participants the opportunity to raise these contextual needs - and they did so.} Model explanations were commonly requested, especially amongst the non-ML experts. Participants with an ML background also wanted to view additional diagnostic plots to assess model performance, which they felt would help them better understand and contextualize the effects of their actions: \textit{``I would love to be able to do this [action] and have the model spit back out, what’s the AUC looks like'' }[ML:04]. Non-ML experts did not express interest in seeing diagnostic plots. Interestingly, there was disagreement amongst participants towards the value of model diagnostic and performance metrics, with one participant summarizing this tension succinctly: \textit{ ``I am little confused if what we’re trying to do is see how good the model trains vs how well the model represents reality [...] you can have an accurate model or a representative model''}[ML:06]. Prior work in interactive ML systems also emphasizes the importance of building representative, over strictly accurate, models~\cite{Bansal:beyond_accuracy:2019}. 


\vspace{1mm}
\noindent\textbf{Concerns about Injecting Bias.}\label{bias-concerns}
Three participants (two non-ML experts and one ML expert) articulated strong concerns about injecting bias into the ML model as a primary reason they would hesitate to engage with a visualization to refine an ML model. For one non-ML expert, it was a reason not to engage at all -- even if they expressed disagreement with the model results, they would still refrain from interacting due to concerns of bias: \textit{``maybe I really disagreed with it[ the classification ] but it is my own view imposed on something cultural [a movie]''}[non-ML:08]. One non-ML expert also doubts their expertise to introduce model refining changes but would feel more confident if they knew safeguards existed: \textit{``I really think that human safeguard with the expertise is a necessary step''}[non-ML:03]. Participants also indicated that the model refinements should be considered experimental and were not to be applied to production settings. 
Overall, these findings show that participants are aware of their potential to introduce bias and this impacts their willingness to steer the model. \rev{While additional components to the IML systems (i.e., model explanations) could help address these concerns, participants' comments also point to the importance of socio-technical structures that include peer safeguards as being important to them. }

\subsubsection{Learning from interactions}\label{interact-learning}
As participants interacted with visual encodings, the study administrators initiated responses to their actions. We asked participants what their expectations were toward what the model was learning, how often it should update, and how they interpreted the responses. 

\vspace{1mm}
\noindent\textbf{Frequency of Model Updates.} 
Participants wanted model feedback in real-time, but they wanted to control the frequency of updates and to only `commit' to certain changes only once they were satisfied.  This form of visual speculative execution~\cite{sperrle2018speculative} aligns with the refine-forecast model update modality described by Strobelt~\textit{et al.}~\cite{strobelt:refineforecast:2022}.

\vspace{1mm}
\noindent\textbf{Learn from Minor Examples, not Major Refinements.} Participants expected the ML model to learn from a small number of interactions. Participants consistently interacted with the marks of movies they were more familiar with than others; they did not express the desire to make changes to those they were unfamiliar with. Additionally, all participants articulated prioritizing their interactions and placing different emphasis on different types of errors: 1) instances where the model was outright wrong (i.e., predicted that \textit{Dr.~Strangelove} is a romance); 2) instances where the model was not completely wrong but should be nudged in a different direction (i.e., whether \textit{Die Hard} is an action or holiday movie); 3) instances where the participant disagreed with both the predicted and labeled genre and wanted to provide their own.  Personal notions of how incorrect the model was, and the `extent of error', also dictated how much weight the model should give these modifications, with error correction given a higher weight than nudges; we discuss this observation in the next subsection (Section~\ref{subsec:uncertainty}).

\rev{It is unlikely that even using their own dataset a participant is intimately familiar with each data point. We argue that participants will always have some prioritization strategy based on familiarity with some subset of the data over others.} It is also important to consider that datasets are frequently reused both in research and in practice~\cite{koch:rrr:2021}. It may be both interesting and relevant to capture \textit{which} subsets are explored and by whom. Interestingly, two participants astutely observed that many other people view and interact with these visualizations, and they wanted a way to see the changes others had made.

\noindent\textbf{Incorporating Uncertainty to Interaction Inputs.}\label{subsec:uncertainty} 
\rev{Many participant concerns and hesitations toward interacting with data encoding marks of visualizations can be summarized by their desire to add uncertainty to their interactions with the model.} Nearly all participants articulated that different types of interactions allowed them to convey a different amount of uncertainty.  For example, changing the size of the mark, its color gradient, or transparency were seen as imprecise interactions meant to convey a larger sense of uncertainty. Conversely, overwriting a value or changing a mark's position was meant to convey less uncertainty. Participants had different preferences for conveying uncertainty. Some prefer less precision:  \textit{``feel that the confidence is more easily visualized here [...] a circle is much easier to say ‘Oh I want to make it bigger or smaller compared to a number where I have to guess some random number [to change it to]''} [non-ML:06].  While others wanted more precision and control: \textit{``I would want to move the points individually, there could be more precision like that [...] because then I can have more control over how confident I am reclassifying something.'}'[non-ML:04]. The choice of visual encoding did impact a participant's ability and willingness to convey a notion of uncertainty. Data targets that participants acted on in other visual encodings were not targeted in others:
\vspace{1mm}
\begin{quote}
    \textit{Admin: In the previous vis you changed the confidence via the [mark] size, do you want to do that in the table?}
    
    \vspace{2mm}
    \textit{non-ML:07: No (emphatic) and I don’t know why. I mean I guess because it’s numbers, don’t ever change the numbers.}
\end{quote}

Prior work on semantic interactions had considered human inputs to be a kind of `soft data'~\cite{Endert:2012:SemanticOG} that the model must calibrate against what is learned from the training data. However, we find that participants also want to convey uncertainty more explicitly through their interactions than the notion of `soft data' accommodates. Kim~\cite{Kim2015InteractiveAI} introduced the concept of an \textit{interacted latent variable} that more closely approximates participants' expectations. Interacted latent variables explicitly include the input uncertainty from the human to calibrate human inputs against what is learned from the data. In applications of a Bayesian Case Model, Kim~\cite{Kim2015InteractiveAI} asks participants to explicitly provide their level of certainty via an input widget, but future work could explore inferring uncertainty directly from interactions.

%
%

\subsection{RQ3: Impacts of ML Expertise}\label{ml-expertise}
Finally, we consider how ML and non-ML experts differed in their interactions and perceptions. Individual differences aside, non-ML experts, in particular, saw the benefit of using a visual and interactive ML system to help them collaborate with their ML expert peers. This perspective from non-ML experts impacted what they wanted out of an IML system.

\vspace{1mm}
\noindent\textbf{ML Experts and not Substitutes for non-ML Expertise.} The preceding results have already alluded to some of the commonalities and differences between ML and non-ML experts. Here we summarize the key points. If we consider only the set of target-interaction pairs (\autoref{fig:res_details}) we see an overlap between the interactions ML and non-ML participants would undertake. We also see that both groups would interact with similar data targets (predicted genre, and probability). While participants from both groups would also introduce new data targets (i.e., Multiclass Genres), in our study, ML experts introduced more (data weights, similarity) compared to non-ML experts.  The most evident difference was toward additional information participants expressed they needed to reduce cognitive load. Non-ML experts preferred explanations of the model's behavior, whereas ML experts wanted more performance metrics (Section~\ref{interaction-hesitancy}). Both groups felt this additional information would inform their interactions and participants assess their effects.  While both ML and non-ML experts at times declined to engage with the visual design probes, their reasoning was different (Section~\ref{no-interaction}). 
However, personal preferences were more indicative of whether, and how, participants interacted with the visual encodings. 
In summary, our findings support prior research~\cite{Qian:2018:MTeaching}. that, despite similarities, ML experts are not a sufficient proxy for non-ML experts when developing IML systems. 

\vspace{1mm}
\noindent\textbf{Working Collaboratively with Others}\label{subsect:explore_commit}
Several non-ML experts also expressed the desire to comment on specific changes, either as a reference for themselves or with ML experts they collaborated with.  One participant indicated they would use interactive model refinement to provide feedback to the ML experts on their team: \textit{``I would love to prototype models because that would let me refine my own notes and recommendations to the person that is potentially designing the model''}[non-ML:03]. 
Prior research~\cite{Wang:2019,sant:automl:2021,Crisan:2021:automl} has shown that such collaborative modes of interaction were desired but largely absent from existing tools. 
Moreover, recent studies show both the necessity and value of challenging the so-called ground-truth of ML data~\cite{denton2021ground} by enabling wider collaboration between data workers~\cite{gordon:jury:2022,davani:disagreements:2021}; our study suggests that interactions could facilitate this collaboration.

\vspace{-2mm}
\subsection{Summary of Findings}
Our study elicited 43 unique interactions that define a set of data target and interaction pairs (\autoref{fig:res_details}). Through these interactions, participants not only modified existing data, but felt that they could introduce new data (class genres, similarity, and data weights) to steer a classification model. We were also able to capture a myriad of considerations that guided participants' interaction - including hesitating or refining from interacting. 
We contextualize our findings between ML and non-ML experts, showcasing similarities and differences between these groups. Overall, our findings both align with and extend interactions discovered in prior work (i.e.,~\cite{Saket:2020:Direct_Manipulation,Bae:iClustering:2020,Drucker:iCluster:2011}), which demonstrates the utility and value of data and design probes for developing visual and interactive  ML systems.

%% file: tex/discussion_conclusion.tex
\section{Discussion}\label{sec:discuss}
\rev{\revMR{Data and design probes provide a canvas for participants and researchers to co-create elements of an interactive system~\cite{Sanders:Probes:2014}.}
Gathering empirical data prior to the development and implementation of IML systems, including those used for visual analysis, is critical to aligning the design of these systems with the needs and expectations of end-users. }
\revMR{Our research and its findings have explored the use of data and visual design probes for conducting pre-design empirical research toward the creation of visual and interactive machine learning systems. 
Collectively, our probes and elicitation study revealed that users consistently target specific types of data with a varied, yet consistent, set of actions (\autoref{fig:res_details}).
We demonstrate that probes are capable of producing tangible artifacts, from participant utterances to their design choices, and we suggest that these can be used to stimulate a more informed discussion around the design and development of visualizations for IML systems. Furthermore, as ML and AI systems grow in complexity and implicate a wider group of stakeholders, it becomes even more important to align these systems with human needs and expectations. Our findings also show that probes are useful tools for teasing apart both differences and commonalities across the continuum of ML/DS expertise. We now reflect on the implications of our findings for developing visual and interactive machine learning systems.}


\subsection{\revMR{Designing Visual \& Interactive ML Systems}}\label{disscussion:design}
\revMR{Our initial goal was to use data and design probes as a precursor to developing an IML system. However, the complexity of the interaction design space and context for these interactions gave us pause. Our findings suggested challenges and limitations of ML models and the incorporation of human refinement that impacts IML system design. While data and design probes do not resolve these issues, they appear to surface them and provide an empirical basis for discussing trade-offs of different approaches without prematurely committing to design choices}

\vspace{1mm}
\noindent\textbf{\revMR{Human Goals and IML Brittleness.}} \revMR{What surprised us about our findings was the tensions they surfaced between what people wanted to do and what present ML models are capable of. As an example, participants wanted to exercise control over both the class label and classification probability; Die Hard should be both a holiday and action movie but with a higher prediction probability on the action genre.
Accommodating this goal is not straightforward. Should an ML model be refined to reflect an end-users expressed uncertainty toward specific data points and classes? Should the initial classification problem be flipped to be treated like a regression problem? 
Participants also added bespoke classes of movies, which introduce out-of-distribution issues should they wish to use their interactively refined model for further prediction. As one participant indicated (see Section~\ref{interact-learning}), the desired features for a system tend to default to a model-centric perspective, focusing on accuracy performance, and do not accommodate human-centric desires like the representativeness of data.}

\vspace{1mm}
\noindent{\textbf{\revMR{Learning from Human Refinement.}}} \revMR{Refining machine learning models through human feedback via visual encodings have been explored using both reinforcement~\cite{el-assady_progressive_2018} and active learning~\cite{Bernard:VIAL:2018}. However, these studies present users with a fixed set of visual encodings and do not ask the critical question: do users want to interact? The answer from our study appears to be that it depends. Our findings suggest limitations in the amount of feedback participants wish to provide (Section~\ref{interact-learning} and Section~\ref{subsec:data_targets}), are impacted by the choice of visual encoding  and their ability to understand them (Section~\ref{subsec:encoding_effects}), their comfort initiating such interactions (Section~\ref{subsec:encoding_effects}), and finally, their preferences to refine ML models via code not by interacting with visual encodings (Section~\ref{subsec:interactions}.) While reinforcement learning from human feedback (RLHF) is showing promise for large language models~\cite{stiennon:rlhf:2022}, the efficacy of such an approach for other types of data is under-explored as is the medium of interaction (e.g., text prompts vs direct manipulation). }   

\revMI{Our study may also have unintentionally captured refinement by voice because study administrators prompted participants to discuss the rationale, which the Wizard could also use to craft a response. Our study focused participants' attention on direct manipulations of visual encodings, thus the effect of voice was not directly under study and likely incidental. However, exploring such multi-modal interactions could be an interesting trajectory of future work.} 

\vspace{1mm}
\noindent{\textbf{\revMR{Providing Feedback to Humans}}}. \revMR{Finally, an often overlooked dimension is communicating to people what the effects of their actions were. We left this consideration entirely open to participants to dictate, both with respect to the frequency (Section~\ref{interact-learning}) and type (Section~\ref{ml-expertise}) of feedback. Here we found that expertise dictated what participants expected to see, with those receiving more formal training expecting to see typical performance metrics while non-experts preferred explanations. However, what was more surprising was that the people also wanted \textit{feedback from other humans} as part of the refinement process. Our results suggest two modes for an IML system. An individual mode that allows analysts of varying expertise to prototype and refine ML models and a complementary feedback mode that allows participants to collaborate with others in the refinement process. Per our results, the feedback mode may extend further to also show how different individuals interact with or annotate data encoding marks.}


\subsection{Limitations}
\rev{The design choices and implementation of our probes as well as our study population and size are limitations of our research. 
While we made these deliberate choices with the intent to minimize the number of assumptions our approach made toward user needs and expectations, we also acknowledge the effects these choices have on our findings.  We chose to develop a minimal set of probes, that made the least assumptions about the components of an IML interface, instead eliciting these aspects from participants and the ML model. For purposes of feasibility, we also limited the number of data points in our study to fifty. While we provide a rationale for these and other choices, there are aspects of the interactive experience that our probes maybe not be able to capture. Scaling design probes, without compromising the value of free expression and creativity that these probes are intended to provide, is a fruitful topic for future work.} \revMR{These design choices may limit some of the generative power of the probes to elicit novel encodings and interactions. The probes are best suited to identify the alignment between targets and actions; the brittleness around this combination may stimulate discussion of the need for novel encodings and interactions, but are not guaranteed to derive them. } \revMI{To evaluate the trade-off between scalability and free expression of our research approach, we would need to implement a system. However, for the reasons described in Section 6.1, this too would have limitations for a proper assessment. We speculate that it may require several different systems that experiment with approaches to human refinement to fully evaluate these trade-offs.}  Lastly, the study size of 20 participants, while commensurate with visualization research on interactive machine learning and elicitation studies~\cite{hohman:gamut:2019,Saket:2020:Direct_Manipulation,Drucker:iCluster:2011,wall:podium:2018,el-assady_progressive_2018,hogan:2016:elicitation}, may not to capture the total spectrum of non-ML and ML expert perspectives. While these limitations are unlikely to hamper the value and utility of data and design probes for visualization research, they do provide an outline for the reasonable scope of our approach.

\subsection{Future Work}
There are several fruitful avenues for future work that expand on our research methodology and its findings. \rev{An immediate next step is to use the findings of our elicitation study and implement a visual and interactive machine learning system. However, our probes as point to many interesting challenges that cannot be solved by the implementation of a single system \revMR{(Section~6.1)} and that remain open research problems for future work.} One interesting area of future work is to expand the definition of so-called ``soft data''~\cite{Endert:2012:SemanticOG}, by allowing people to add their own perceptions of uncertainty \textit{into} model feedback. Prior work by Kim~\cite{Kim2015InteractiveAI} provides an interesting avenue via interacted latent variables, which could be explored concomitantly with visualization research in Bayesian modeling~\cite{Kim:2019:Bayescog,Kim:2021:BayesAssist} and uncertainty~\cite{padilla:2020:uncertainty,Pu:202:GGP}. Another interesting area of future work is to leverage semantic interactions to investigate how people construct and label datasets. For non-ML experts, or perhaps even the lay public more generally, interacting through visualizations may be a more accessible way to communicate their own personal notion of a ``ground truth''~\cite{denton2021ground} and even elicit richer contextual information than current approaches afford~\cite{Bernard:VIAL:2018,Russakovsky_2015_CVPR,Chang:2017:Labels,gordon:jury:2022,davani:disagreements:2021}. Finally, we see promise in further exploring the ways that semantic interactions can facilitate collaboration amongst teams of data workers and especially between ML and non-ML experts. This team collaboration appears to add safeguards to the processes of updating ML models as a greater number of non-ML experts begin to undertake this task. Of interest would be to examine whether semantic interactions could enable an asynchronous and multi-user collaboration around model refinement. 

\section{Conclusions}
\rev{Visual and interactive machine learning systems can provide avenues for data analysts with varied ML expertise to steer ML models with human guidance. However, designing such systems is uniquely hard and may make many assumptions about the needs and expectations of end-users. We examined the use of data and design probes to elicit how participants would interact with visual encodings, via direct manipulation, to update an ML model. Our findings show that the use of probes can not only capture a design space of interactions but also participants' perceptions and hesitations toward interacting with an ML model. Overall, our study demonstrates the value of expanding the visualization research toolbox to include data and design probes as a way of gathering formative empirical evidence for developing IML systems.}

%% file: main.bbl
\begin{thebibliography}{10}
\providecommand{\url}[1]{#1}
\csname url@samestyle\endcsname
\providecommand{\newblock}{\relax}
\providecommand{\bibinfo}[2]{#2}
\providecommand{\BIBentrySTDinterwordspacing}{\spaceskip=0pt\relax}
\providecommand{\BIBentryALTinterwordstretchfactor}{4}
\providecommand{\BIBentryALTinterwordspacing}{\spaceskip=\fontdimen2\font plus
\BIBentryALTinterwordstretchfactor\fontdimen3\font minus
  \fontdimen4\font\relax}
\providecommand{\BIBforeignlanguage}[2]{{%
\expandafter\ifx\csname l@#1\endcsname\relax
\typeout{** WARNING: IEEEtran.bst: No hyphenation pattern has been}%
\typeout{** loaded for the language `#1'. Using the pattern for}%
\typeout{** the default language instead.}%
\else
\language=\csname l@#1\endcsname
\fi
#2}}
\providecommand{\BIBdecl}{\relax}
\BIBdecl

\bibitem{collins:2018:guidance}
\BIBentryALTinterwordspacing
C.~Collins, N.~Andrienko, T.~Schreck, J.~Yang, J.~Choo, U.~Engelke, A.~Jena,
  and T.~Dwyer, ``Guidance in the human–machine analytics process,''
  \emph{Visual Informatics}, vol.~2, no.~3, pp. 166--180, 2018. [Online].
  Available:
  \url{https://www.sciencedirect.com/science/article/pii/S2468502X1830041X}
\BIBentrySTDinterwordspacing

\bibitem{sacha:2017:wywc}
D.~Sacha, M.~Sedlmair, L.~Zhang, J.~A. Lee, J.~Peltonen, D.~Weiskopf, S.~C.
  North, and D.~A. Keim, ``What you see is what you can change: Human-centered
  machine learning by interactive visualization,'' \emph{Neurocomputing}, vol.
  268, pp. 164--175, 2017.

\bibitem{SPERRLE:coadative:2021}
F.~Sperrle, A.~Jeitler, J.~Bernard, D.~Keim, and M.~El-Assady, ``Co-adaptive
  visual data analysis and guidance processes,'' \emph{Computers \& Graphics},
  2021.

\bibitem{Sperrle:HCE_STAR:2021}
F.~Sperrle, M.~El-Assady, G.~Guo, R.~Borgo, D.~H. Chau, A.~Endert, and D.~Keim,
  ``A survey of human-centered evaluations in human-centered machine
  learning,'' \emph{Computer Graphics Forum}, vol.~40, no.~3, pp. 543--567,
  2021.

\bibitem{Boukhelifa:eval_attribution:2020}
N.~Boukhelifa, A.~Bezerianos, R.~Chang, C.~Collins, S.~Drucker, A.~Endert,
  J.~Hullman, C.~North, and M.~Sedlmair, ``Challenges in evaluating interactive
  visual machine learning systems,'' \emph{IEEE Computer Graphics and
  Applications}, vol.~40, no.~6, pp. 88--96, 2020.

\bibitem{dudley:UI_IML:2018}
\BIBentryALTinterwordspacing
J.~J. Dudley and P.~O. Kristensson, ``A review of user interface design for
  interactive machine learning,'' \emph{ACM Trans. Interact. Intell. Syst.},
  vol.~8, no.~2, jun 2018. [Online]. Available:
  \url{https://doi.org/10.1145/3185517}
\BIBentrySTDinterwordspacing

\bibitem{Combemale:DACI:2021}
B.~Combemale, J.~Kienzle, G.~Mussbacher, H.~Ali, D.~Amyot, M.~Bagherzadeh,
  E.~Batot, N.~Bencomo, B.~Benni, J.-M. Bruel, J.~Cabot, B.~H. Cheng,
  P.~Collet, G.~Engels, R.~Heinrich, J.-M. Jezequel, A.~Koziolek, S.~Mosser,
  R.~Reussner, H.~Sahraoui, R.~Saini, J.~Sallou, S.~Stinckwich, E.~Syriani, and
  M.~Wimmer, ``A hitchhiker's guide to model-driven engineering for
  data-centric systems,'' \emph{IEEE Software}, vol.~38, no.~4, pp. 71--84,
  2021.

\bibitem{Subramonyam:DACI:2021}
\BIBentryALTinterwordspacing
H.~Subramonyam, C.~Seifert, and E.~Adar, ``How can human-centered design shape
  data-centric ai?'' 2021. [Online]. Available:
  \url{https://sites.google.com/view/hcai-human-centered-ai-neurips/home}
\BIBentrySTDinterwordspacing

\bibitem{Sambasivan:DataWork:2021}
N.~Sambasivan, S.~Kapania, H.~Highfill, D.~Akrong, P.~Paritosh, and L.~M.
  Aroyo, ``“everyone wants to do the model work, not the data work”: Data
  cascades in high-stakes ai,'' in \emph{Proc. CHI'21s}, New York, NY, USA,
  2021.

\bibitem{Ender:beyond_controlpanels:2013}
A.~Endert, L.~Bradel, and C.~North, ``Beyond control panels: Direct
  manipulation for visual analytics,'' \emph{IEEE Computer Graphics and
  Applications}, vol.~33, no.~4, pp. 6--13, 2013.

\bibitem{Qian:2018:MTeaching}
Q.~Yang, J.~Suh, N.-C. Chen, and G.~Ramos, ``Grounding interactive machine
  learning tool design in how non-experts actually build models,'' in
  \emph{Proc. DIS '18}, New York, NY, USA, 2018, p. 573–584.

\bibitem{Swati:design_nonexperts:2021}
S.~Mishra and J.~M. Rzeszotarski, ``Designing interactive transfer learning
  tools for ml non-experts,'' in \emph{Proc. CHI'21}, 2021.

\bibitem{Yang:HAI_hard:2020}
Q.~Yang, A.~Steinfeld, C.~Ros\'{e}, and J.~Zimmerman, ``Re-examining whether,
  why, and how human-ai interaction is uniquely difficult to design,'' in
  \emph{Proc. CHI'20}, 2020, p. 1–13.

\bibitem{Sanders:Probes:2014}
\BIBentryALTinterwordspacing
E.~B.-N. Sanders and P.~J. Stappers, ``Probes, toolkits and prototypes: three
  approaches to making in codesigning,'' \emph{CoDesign}, vol.~10, no.~1, pp.
  5--14, 2014. [Online]. Available:
  \url{https://doi.org/10.1080/15710882.2014.888183}
\BIBentrySTDinterwordspacing

\bibitem{Hutchins1985DirectMI}
E.~L. Hutchins, J.~D. Hollan, and D.~A. Norman, ``Direct manipulation
  interfaces,'' \emph{Hum. Comput. Interact.}, vol.~1, pp. 311--338, 1985.

\bibitem{Graham:probes:2008}
C.~Graham and M.~Rouncefield, ``Probes and participation,'' in \emph{Proc.
  PDC'08}, 2008, p. 194–197.

\bibitem{Subramonyam:Prtotyping:2021}
H.~Subramonyam, C.~Seifert, and E.~Adar, \emph{ProtoAI: Model-Informed
  Prototyping for AI-Powered Interfaces}, 2021, p. 48–58.

\bibitem{Lee2019AHP}
\BIBentryALTinterwordspacing
D.~Lee, S.~Macke, D.~Xin, A.~Lee, S.~Huang, and A.~G. Parameswaran, ``A
  human-in-the-loop perspective on automl: Milestones and the road ahead,''
  \emph{IEEE Data Eng. Bull.}, vol.~42, no.~2, pp. 59--70, 2019. [Online].
  Available: \url{http://sites.computer.org/debull/A19june/p59.pdf}
\BIBentrySTDinterwordspacing

\bibitem{Zanzotto:HAI:2019}
F.~M. Zanzotto, ``Viewpoint: Human-in-the-loop artificial intelligence,''
  \emph{Journal of Artificial Intelligence Research}, vol.~64, p. 243–252,
  Feb 2019.

\bibitem{Crisan:2021:automl}
A.~Crisan and B.~Fiore-Gartland, \emph{Fits and Starts: Enterprise Use of
  AutoML and the Role of Humans in the Loop}, 2021.

\bibitem{dellermann:future_collab:2021}
\BIBentryALTinterwordspacing
D.~Dellermann, A.~Calma, N.~Lipusch, T.~Weber, S.~Weigel, and P.~Ebel, ``The
  future of human-ai collaboration: a taxonomy of design knowledge for hybrid
  intelligence systems,'' 2021. [Online]. Available:
  \url{https://arxiv.org/abs/2105.03354}
\BIBentrySTDinterwordspacing

\bibitem{Ben:2020:HCAI}
B.~Shneiderman, ``Human-centered artificial intelligence: Reliable, safe \&
  trustworthy,'' \emph{International Journal of Human–Computer Interaction},
  vol.~36, no.~6, pp. 495--504, 2020.

\bibitem{Gil:HGML:2019}
Y.~Gil, J.~Honaker, S.~Gupta, Y.~Ma, V.~D'Orazio, D.~Garijo, S.~Gadewar,
  Q.~Yang, and N.~Jahanshad, ``Towards human-guided machine learning,'' in
  \emph{Proc. IUI'19}, 2019, p. 614–624.

\bibitem{Ceneda:2017:guidance}
D.~Ceneda, T.~Gschwandtner, T.~May, S.~Miksch, H.-J. Schulz, M.~Streit, and
  C.~Tominski, ``Characterizing guidance in visual analytics,'' \emph{IEEE
  Transactions on Visualization and Computer Graphics}, vol.~23, no.~1, pp.
  111--120, 2017.

\bibitem{Wijk:2005}
J.~J. van Wijk, ``The value of visualization,'' in \emph{VIS 05. IEEE
  Visualization, 2005.}, 2005, pp. 79--86.

\bibitem{Kindlman:2014}
G.~Kindlmann and C.~Scheidegger, ``An algebraic process for visualization
  design,'' \emph{IEEE Transactions on Visualization and Computer Graphics},
  vol.~20, no.~12, pp. 2181--2190, 2014.

\bibitem{McNutt:2020:CHI}
\BIBentryALTinterwordspacing
A.~McNutt, G.~Kindlmann, and M.~Correll, ``Surfacing visualization mirages,''
  in \emph{Proc CHI'20}.\hskip 1em plus 0.5em minus 0.4em\relax New York, NY,
  USA: Association for Computing Machinery, 2020, p. 1–16. [Online].
  Available: \url{https://doi.org/10.1145/3313831.3376420}
\BIBentrySTDinterwordspacing

\bibitem{ben:DM:1983}
B.~Shneiderman, ``Direct manipulation: A step beyond programming languages,''
  \emph{Computer}, vol.~16, no.~8, pp. 57--69, 1983.

\bibitem{Jiang2019RecentRA}
L.~Jiang, S.~Liu, and C.~Chen, ``Recent research advances on interactive
  machine learning,'' \emph{Journal of Visualization}, vol.~22, pp. 401--417,
  2019.

\bibitem{Endert2017}
A.~Endert, W.~Ribarsky, C.~Turkay, B.~L.~W. Wong, I.~Nabney, I.~D. Blanco, and
  F.~Rossi, ``The state of the art in integrating machine learning into visual
  analytics,'' \emph{Computer Graphics Forum}, vol.~36, no.~8, pp. 458--486,
  2017.

\bibitem{ramos2020interactive}
G.~Ramos, C.~Meek, P.~Simard, J.~Suh, and S.~Ghorashi, ``Interactive machine
  teaching: a human-centered approach to building machine-learned models,''
  \emph{Human–Computer Interaction}, vol.~35, no. 5-6, pp. 413--451, April
  2020.

\bibitem{Holstein:practitioner_fairness:2019}
\BIBentryALTinterwordspacing
K.~Holstein, J.~Wortman~Vaughan, H.~Daum\'{e}, M.~Dudik, and H.~Wallach,
  ``Improving fairness in machine learning systems: What do industry
  practitioners need?'' in \emph{Proc. CHI'19}, 2019, p. 1–16. [Online].
  Available: \url{https://doi.org/10.1145/3290605.3300830}
\BIBentrySTDinterwordspacing

\bibitem{Endert:2012:SemanticOG}
A.~Endert, P.~Fiaux, and C.~North, ``Semantic interaction for sensemaking:
  Inferring analytical reasoning for model steering,'' \emph{IEEE Transactions
  on Visualization and Computer Graphics}, vol.~18, no.~12, pp. 2879--2888,
  2012.

\bibitem{Bradel:Starspire:1995}
L.~Bradel, C.~North, L.~House, and S.~Leman, ``Multi-model semantic interaction
  for text analytics,'' in \emph{Proc IEEE VAST'14}, 2014, pp. 163--172.

\bibitem{Lee:iVisClustering:2012}
H.~Lee, J.~Kihm, J.~Choo, J.~T. Stasko, and H.~Park, ``ivisclustering: An
  interactive visual document clustering via topic modeling,'' \emph{Computer
  Graphics Forum}, vol.~31, 2012.

\bibitem{Choo:Utopian:2013}
J.~Choo, C.~Lee, C.~K. Reddy, and H.~Park, ``Utopian: User-driven topic
  modeling based on interactive nonnegative matrix factorization,'' \emph{IEEE
  Transactions on Visualization and Computer Graphics}, vol.~19, no.~12, pp.
  1992--2001, 2013.

\bibitem{Demiral:2014}
{\c{C}}.~Demiralp, M.~S. Bernstein, and J.~Heer, ``Learning perceptual kernels
  for visualization design,'' \emph{IEEE Transactions on Visualization and
  Computer Graphics}, vol.~20, no.~12, pp. 1933--1942, 2014.

\bibitem{Saket:2020:Direct_Manipulation}
B.~{Saket}, S.~{Huron}, C.~{Perin}, and A.~{Endert}, ``Investigating direct
  manipulation of graphical encodings as a method for user interaction,''
  \emph{IEEE Transactions on Visualization and Computer Graphics}, vol.~26,
  no.~1, pp. 482--491, 2020.

\bibitem{Saket:visdemo:2017}
B.~Saket, H.~Kim, E.~T. Brown, and A.~Endert, ``Visualization by demonstration:
  An interaction paradigm for visual data exploration,'' \emph{IEEE
  Transactions on Visualization and Computer Graphics}, vol.~23, no.~1, pp.
  331--340, 2017.

\bibitem{Hartmann:SensorDM:2007}
\BIBentryALTinterwordspacing
B.~Hartmann, L.~Abdulla, M.~Mittal, and S.~R. Klemmer, ``Authoring sensor-based
  interactions by demonstration with direct manipulation and pattern
  recognition,'' in \emph{Proc. CHI '07}.\hskip 1em plus 0.5em minus
  0.4em\relax New York, NY, USA: Association for Computing Machinery, 2007, p.
  145–154. [Online]. Available: \url{https://doi.org/10.1145/1240624.1240646}
\BIBentrySTDinterwordspacing

\bibitem{Brown:2012}
E.~T. Brown, J.~Lie, C.~E. Brodely, and R.~Chang, ``Dis-function: Learning
  distance functions interactively,'' in \emph{2012 IEEE Conference on Visual
  Analytics Science and Technology (VAST)}, 2012, pp. 83--92.

\bibitem{wall:podium:2018}
E.~Wall, S.~Das, R.~Chawla, B.~Kalidindi, E.~T. Brown, and A.~Endert, ``Podium:
  Ranking data using mixed-initiative visual analytics,'' \emph{IEEE
  Transactions on Visualization and Computer Graphics}, vol.~24, no.~1, pp.
  288--297, 2018.

\bibitem{Gehrmann:MLhooks:2020}
S.~Gehrmann, H.~Strobelt, R.~Krüger, H.~Pfister, and A.~M. Rush, ``Visual
  interaction with deep learning models through collaborative semantic
  inference,'' \emph{IEEE Transactions on Visualization and Computer Graphics},
  vol.~26, no.~1, pp. 884--894, 2020.

\bibitem{Drucker:iCluster:2011}
S.~M. Drucker, D.~Fisher, and S.~Basu, ``Helping users sort faster with
  adaptive machine learning recommendations,'' in \emph{Proc. INTERACT'11},
  2011, p. 187–203.

\bibitem{Bernard:VIAL:2018}
J.~Bernard, M.~Zeppelzauer, M.~Sedlmair, and W.~Aigner, ``Vial: a unified
  process for visual interactive labeling,'' \emph{The Visual Computer},
  vol.~34, no.~9, pp. 1189--1207, Sep 2018.

\bibitem{Bernard:2018}
J.~Bernard, M.~Hutter, M.~Zeppelzauer, D.~Fellner, and M.~Sedlmair, ``Comparing
  visual-interactive labeling with active learning: An experimental study,''
  \emph{IEEE Transactions on Visualization and Computer Graphics}, vol.~24,
  no.~1, pp. 298--308, 2018.

\bibitem{hohman:data_iteration:2020}
\BIBentryALTinterwordspacing
F.~Hohman, K.~Wongsuphasawat, M.~B. Kery, and K.~Patel, ``Understanding and
  visualizing data iteration in machine learning,'' in \emph{Proc. CHI'20},
  2020, p. 1–13. [Online]. Available:
  \url{https://doi.org/10.1145/3313831.3376177}
\BIBentrySTDinterwordspacing

\bibitem{Sacha:ReviewDIM:2017}
D.~Sacha, L.~Zhang, M.~Sedlmair, J.~A. Lee, J.~Peltonen, D.~Weiskopf, S.~C.
  North, and D.~A. Keim, ``Visual interaction with dimensionality reduction: A
  structured literature analysis,'' \emph{IEEE Transactions on Visualization
  and Computer Graphics}, vol.~23, no.~1, pp. 241--250, 2017.

\bibitem{Yuan:survey:2021}
J.~Yuan, C.~Chen, W.~Yang, M.~Liu, J.~Xia, and S.~Liu, ``A survey of visual
  analytics techniques for machine learning,'' \emph{Computational Visual
  Media}, vol.~7, no.~1, pp. 3--36, Mar 2021.

\bibitem{Bae:iClustering:2020}
J.~Bae, T.~Helldin, M.~Riveiro, S.~Nowaczyk, M.-R. Bouguelia, and G.~Falkman,
  ``Interactive clustering: A comprehensive review,'' \emph{ACM Comput. Surv.},
  vol.~53, no.~1, feb 2020.

\bibitem{Hutchinson:techprobe:2003}
H.~Hutchinson, W.~Mackay, B.~Westerlund, B.~B. Bederson, A.~Druin, C.~Plaisant,
  M.~Beaudouin-Lafon, S.~Conversy, H.~Evans, H.~Hansen, N.~Roussel, and
  B.~Eiderb\"{a}ck, ``Technology probes: Inspiring design for and with
  families,'' in \emph{Proc. CHI'03}.\hskip 1em plus 0.5em minus 0.4em\relax
  Association for Computing Machinery, 2003, p. 17–24.

\bibitem{Lam:Scenarios:2012}
H.~Lam, E.~Bertini, P.~Isenberg, C.~Plaisant, and S.~Carpendale, ``Empirical
  studies in information visualization: Seven scenarios,'' \emph{IEEE
  Transactions on Visualization and Computer Graphics}, vol.~18, no.~9, pp.
  1520--1536, 2012.

\bibitem{Broadley:VisualisingHD:2013}
C.~Broadley, ``Visualising human-centred design relationships : a toolkit for
  participation,'' 2013.

\bibitem{Stumpf:InteractingMW:2009}
S.~Stumpf, V.~Rajaram, L.~Li, W.-K. Wong, M.~Burnett, T.~Dietterich,
  E.~Sullivan, and J.~Herlocker, ``Interacting meaningfully with machine
  learning systems: Three experiments,'' \emph{Int. J. Hum. Comput. Stud.},
  vol.~67, no.~8, pp. 639--662, 2009.

\bibitem{Subramonyam:cocreationg:2021}
H.~Subramonyam, C.~Seifert, and E.~Adar, ``Towards a process model for
  co-creating ai experiences,'' in \emph{Proc. DIS'21}, 2021, p. 1529–1543.

\bibitem{browne:ml_wizard_of_oz:2019}
\BIBentryALTinterwordspacing
J.~T. Browne, ``Wizard of oz prototyping for machine learning experiences,'' in
  \emph{Proc. CHI EA '19}.\hskip 1em plus 0.5em minus 0.4em\relax New York, NY,
  USA: Association for Computing Machinery, 2019, p. 1–6. [Online].
  Available: \url{https://doi.org/10.1145/3290607.3312877}
\BIBentrySTDinterwordspacing

\bibitem{Maulsby:Woo:1993}
D.~Maulsby, S.~Greenberg, and R.~Mander, ``Prototyping an intelligent agent
  through wizard of oz,'' in \emph{Proc. CHI '93}, 1993, p. 277–284.

\bibitem{Dove:MLWoo"2017}
G.~Dove, K.~Halskov, J.~Forlizzi, and J.~Zimmerman, ``Ux design innovation:
  Challenges for working with machine learning as a design material,'' in
  \emph{Proc. CHI'17}, 2017.

\bibitem{hohman:gamut:2019}
F.~Hohman, A.~Head, R.~Caruana, R.~DeLine, and S.~M. Drucker, ``Gamut: A design
  probe to understand how data scientists understand machine learning models,''
  in \emph{Proc. CHI'19}.\hskip 1em plus 0.5em minus 0.4em\relax ACM, 2019.

\bibitem{Brehmer:design:2014}
M.~Brehmer, S.~Carpendale, B.~Lee, and M.~Tory, ``Pre-design empiricism for
  information visualization: Scenarios, methods, and challenges,'' in
  \emph{Proc. BELIV '14}, 2014, p. 147–151.

\bibitem{hogan:2016:elicitation}
T.~Hogan, U.~Hinrichs, and E.~Hornecker, ``The elicitation interview technique:
  Capturing people's experiences of data representations,'' \emph{IEEE
  Transactions on Visualization \& Computer Graphics}, vol.~22, no.~12, pp.
  2579--2593, 2016.

\bibitem{Xin2021WhitherAU}
D.~Xin, E.~Y. Wu, D.~Lee, N.~Salehi, and A.~Parameswaran, ``Whither automl?
  understanding the role of automation in machine learning workflows,''
  \emph{ArXiv}, vol. abs/2101.04834, 2021.

\bibitem{Creswell_Poth_2018}
J.~W. Creswell and C.~N. Poth, \emph{Qualitative inquiry \& Research Design:
  Choosing Among Five Approaches}, fourth edition~ed.\hskip 1em plus 0.5em
  minus 0.4em\relax Los Angeles, Calif: Sage Publications, 2018.

\bibitem{valdez:anchoring:2018}
A.~C. Valdez, M.~Ziefle, and M.~Sedlmair, ``Priming and anchoring effects in
  visualization,'' \emph{IEEE Transactions on Visualization and Computer
  Graphics}, vol.~24, no.~1, pp. 584--594, 2018.

\bibitem{Schapire:boosting:2005}
R.~E. Schapire, ``The strength of weak learnability,'' \emph{Machine Learning},
  vol.~5, pp. 197--227, 2005.

\bibitem{el-assady_progressive_2018}
\BIBentryALTinterwordspacing
M.~El-Assady, R.~Sevastjanova, F.~Sperrle, D.~Keim, and C.~Collins,
  ``Progressive {Learning} of {Topic} {Modeling} {Parameters}: {A} {Visual}
  {Analytics} {Framework},'' \emph{IEEE Transactions on Visualization and
  Computer Graphics}, vol.~24, no.~1, pp. 382--391, Jan. 2018. [Online].
  Available: \url{http://ieeexplore.ieee.org/document/8019825/}
\BIBentrySTDinterwordspacing

\bibitem{Crisan:2021:text}
A.~Crisan and M.~Correll, ``User ex machina: Simulation as a design probe in
  human-in-the-loop text analytics,'' in \emph{Proc. CHI'21}, 2021.

\bibitem{Bansal:beyond_accuracy:2019}
\BIBentryALTinterwordspacing
G.~Bansal, B.~Nushi, E.~Kamar, W.~S. Lasecki, D.~S. Weld, and E.~Horvitz,
  ``Beyond accuracy: The role of mental models in human-ai team performance,''
  \emph{Proceedings of the AAAI Conference on Human Computation and
  Crowdsourcing}, vol.~7, no.~1, pp. 2--11, Oct. 2019. [Online]. Available:
  \url{https://ojs.aaai.org/index.php/HCOMP/article/view/5285}
\BIBentrySTDinterwordspacing

\bibitem{sperrle2018speculative}
\BIBentryALTinterwordspacing
F.~Sperrle, J.~Bernard, M.~Sedlmair, D.~Keim, and M.~El-Assady, ``Speculative
  execution for guided visual analytics,'' in \emph{Proc. IEEE VIS Work. Mach.
  Learn. from User Interact. Vis. Anal.}, 2018. [Online]. Available:
  \url{https://learningfromusersworkshop.github.io/papers/SpecEx.pdf}
\BIBentrySTDinterwordspacing

\bibitem{strobelt:refineforecast:2022}
H.~Strobelt, J.~Kinley, R.~Krueger, J.~Beyer, H.~Pfister, and A.~M. Rush,
  ``Genni: Human-ai collaboration for data-backed text generation,'' \emph{IEEE
  Transactions on Visualization and Computer Graphics}, vol.~28, no.~1, pp.
  1106--1116, 2022.

\bibitem{koch:rrr:2021}
B.~Koch, E.~Denton, A.~Hanna, and J.~G. Foster, ``Reduced, reused and recycled:
  The life of a dataset in machine learning research,'' 2021.

\bibitem{Kim2015InteractiveAI}
\BIBentryALTinterwordspacing
B.~Kim, ``Interactive and interpretable machine learning models for human
  machine collaboration,'' 2015. [Online]. Available:
  \url{https://beenkim.github.io/papers/BKimPhDThesis.pdf}
\BIBentrySTDinterwordspacing

\bibitem{Wang:2019}
D.~Wang, J.~D. Weisz, M.~Muller, P.~Ram, W.~Geyer, C.~Dugan, Y.~Tausczik,
  H.~Samulowitz, and A.~Gray, ``Human-ai collaboration in data science:
  Exploring data scientists’ perceptions of automated ai,'' 2019.

\bibitem{sant:automl:2021}
\BIBentryALTinterwordspacing
S.~K. Karmaker, M.~M. Hassan, M.~J. Smith, L.~Xu, C.~Zhai, and
  K.~Veeramachaneni, ``Automl to date and beyond: Challenges and
  opportunities,'' 2021. [Online]. Available:
  \url{https://arxiv.org/abs/2010.10777}
\BIBentrySTDinterwordspacing

\bibitem{denton2021ground}
\BIBentryALTinterwordspacing
E.~Denton, M.~Díaz, I.~Kivlichan, V.~Prabhakaran, and R.~Rosen, ``Whose ground
  truth? accounting for individual and collective identities underlying dataset
  annotation,'' 2021. [Online]. Available:
  \url{https://arxiv.org/abs/2112.04554}
\BIBentrySTDinterwordspacing

\bibitem{gordon:jury:2022}
M.~L. Gordon, M.~S. Lam, J.~S. Park, K.~Patel, J.~T. Hancock, T.~Hashimoto, and
  M.~S. Bernstein, ``Jury learning: Integrating dissenting voices into machine
  learning models,'' 2022.

\bibitem{davani:disagreements:2021}
\BIBentryALTinterwordspacing
A.~M. Davani, M.~Díaz, and V.~Prabhakaran, ``Dealing with disagreements:
  Looking beyond the majority vote in subjective annotations,'' 2021. [Online].
  Available: \url{https://arxiv.org/abs/2110.05719}
\BIBentrySTDinterwordspacing

\bibitem{stiennon:rlhf:2022}
N.~Stiennon, L.~Ouyang, J.~Wu, D.~M. Ziegler, R.~Lowe, C.~Voss, A.~Radford,
  D.~Amodei, and P.~Christiano, ``Learning to summarize from human feedback,''
  2022.

\bibitem{Kim:2019:Bayescog}
\BIBentryALTinterwordspacing
Y.-S. Kim, L.~A. Walls, P.~Krafft, and J.~Hullman, ``A bayesian cognition
  approach to improve data visualization,'' in \emph{Proc CHI'19}.\hskip 1em
  plus 0.5em minus 0.4em\relax New York, NY, USA: Association for Computing
  Machinery, 2019, p. 1–14. [Online]. Available:
  \url{https://doi.org/10.1145/3290605.3300912}
\BIBentrySTDinterwordspacing

\bibitem{Kim:2021:BayesAssist}
Y.-S. Kim, P.~Kayongo, M.~Grunde-McLaughlin, and J.~Hullman,
  ``Bayesian-assisted inference from visualized data,'' \emph{IEEE Transactions
  on Visualization \& Computer Graphics}, vol.~27, no.~02, pp. 989--999, feb
  2021.

\bibitem{padilla:2020:uncertainty}
\BIBentryALTinterwordspacing
L.~Padilla, M.~Kay, and J.~Hullman, ``Uncertainty visualization,'' Apr 2020.
  [Online]. Available: \url{psyarxiv.com/ebd6r}
\BIBentrySTDinterwordspacing

\bibitem{Pu:202:GGP}
X.~Pu and M.~Kay, ``A probabilistic grammar of graphics,'' in \emph{Proc
  CHI'20}.\hskip 1em plus 0.5em minus 0.4em\relax New York, NY, USA:
  Association for Computing Machinery, 2020, p. 1–13.

\bibitem{Russakovsky_2015_CVPR}
O.~Russakovsky, L.-J. Li, and L.~Fei-Fei, ``Best of both worlds: Human-machine
  collaboration for object annotation,'' in \emph{Proc CVPR'15}, June 2015.

\bibitem{Chang:2017:Labels}
J.~C. Chang, S.~Amershi, and E.~Kamar, ``Revolt: Collaborative crowdsourcing
  for labeling machine learning datasets,'' in \emph{Proceedings CHI'17}, ser.
  CHI '17, 2017, p. 2334–2346.

\end{thebibliography}
